\documentclass[preprint,11pt,authoryear]{elsarticle}

\usepackage[utf8]{inputenc}
\usepackage[T1]{fontenc}
\usepackage{booktabs,longtable,array,etoolbox}
\usepackage{amsmath,amssymb,amsfonts}
\usepackage{microtype}
\microtypesetup{expansion=false}
\usepackage{xcolor}
\DeclareRobustCommand{\rev}[1]{#1}
\usepackage{lineno}
\usepackage[colorlinks=true,linkcolor=blue!55!black,
  citecolor=blue!55!black,urlcolor=blue!55!black]{hyperref}
\journal{Transportation Research Part C: Emerging Technologies}
\newcolumntype{P}[1]{>{\raggedright\arraybackslash}p{#1}}
\newcommand{\tabitem}[1]{\par\noindent\hangindent=0.9em\makebox[0.9em][l]{\textbullet}#1}
\AtBeginEnvironment{longtable}{\small}
\newcommand{\mobilityrowrule}{\specialrule{0.25pt}{0.12em}{0.12em}}
\newenvironment{insightbox}{%
  \par\smallskip\noindent\begin{center}\begin{minipage}{0.94\textwidth}%
  \setlength{\parindent}{0pt}\hrule\smallskip%
}{%
  \smallskip\hrule\end{minipage}\end{center}\smallskip%
}

\hypersetup{
  pdftitle={Learning Sequential Mobility Choice: A Review of Route and Activity Choice through Inverse Reinforcement Learning and Imitation Learning},
  pdfauthor={Hung Tran, Viet Bui, Tien Mai},
  pdfsubject={A critical review of sequential mobility choice, route choice, activity choice, inverse reinforcement learning, imitation learning, and inverse constraint learning},
  pdfkeywords={sequential mobility choice, route choice, activity choice, recursive logit, inverse reinforcement learning, constrained MDP, inverse constraint learning, imitation learning, dynamic discrete choice, machine learning}
}

\begin{document}
\begin{frontmatter}

\title{\rev{Learning Sequential Mobility Choice: A Review of Route and Activity Choice through Inverse Reinforcement Learning and Imitation Learning}}

\author[smu]{Hung Tran}
\author[smu]{Viet Bui}
\author[smu]{Tien Mai\corref{cor1}}
\ead{atmai@smu.edu.sg}
\cortext[cor1]{Corresponding author.}
\affiliation[smu]{organization={School of Computing and Information Systems, Singapore Management University},
  addressline={80 Stamford Road},
  postcode={178902},
  city={Singapore},
  country={Singapore}}

\begin{abstract}
Route and activity choice are distinct transportation problems that both require models of feasible decisions unfolding over networks and time. This critical integrative review connects transportation choice modeling with inverse reinforcement learning (IRL) and imitation learning (IL), while distinguishing evidence from transportation applications, transferable methods from other fields, and emerging proposals. We develop a four-layer \emph{sequential mobility choice} framework comprising the environment, behavioral objective, stochastic choice mechanism, and observation process. Under stated assumptions, recursive logit, logit dynamic discrete choice, and maximum-entropy IRL use the same soft Bellman recursion linking future opportunities to current choice probabilities. Expected state--action visitation also satisfies conservation equations analogous to network flows. These mathematical connections do not make utility, reward, policy, occupancy, constraints, and observation error behaviorally interchangeable. Transportation evidence is strongest for network-scale planning, context-dependent reward learning, inference from incomplete trajectories, and activity-schedule generation, but remains limited for actual interventions and transfer across networks. We therefore propose a behaviorally disciplined hybrid architecture that keeps feasible actions, interpretable trade-offs, observation processes, and system feedback explicit while using machine learning for scalable computation, contextual representation, heterogeneity, and data integration.
\end{abstract}


\end{frontmatter}


\section{Introduction}

Travel behavior is rarely an isolated selection from a menu. A driver traverses a sequence of connected roads; a transit passenger combines services and transfers; and a household arranges activities, locations, modes, and durations under time, money, and coordination constraints. Each decision changes the opportunities available later. A useful behavioral model must therefore explain not only which alternative is selected now, but how feasible and purposeful sequences emerge--and how those sequences change when networks, prices, schedules, information, or constraints change.

Transportation modeling and machine learning (ML) have developed different languages for this problem. Random-utility models, recursive route choice, activity-based microsimulation, dynamic discrete choice (DDC), and traffic assignment emphasize utility, substitution, feasibility, equilibrium, and policy interpretation. Markov decision processes (MDPs), inverse reinforcement learning (IRL), imitation learning (IL), offline reinforcement learning (RL), and sequence models emphasize reward, policies, occupancy, representation, and scalable prediction. The two traditions increasingly study the same empirical object--observed trajectories--but often target different estimands and use different validation standards. This gap matters: matching observed paths is not the same as recovering preferences, and a reward that reproduces current behavior need not support a credible toll, closure, or accessibility counterfactual.

We use \emph{sequential mobility choice} for the common model class: a context-conditioned sequence of feasible travel and activity decisions whose consequences accumulate through a network and through time. Route choice is its spatial, usually destination-conditioned instance. Activity choice enlarges the state to include purposes, timing and duration, remaining resources, prior commitments, and household coordination. The common formulation exposes reusable mathematics and learning tools without pretending that the applications are identical. A next-link model can be adequate for routing and still be invalid for activity duration, endogenous destination formation, or coordination among household members.

\begin{insightbox}
\textbf{Core message.} The learned object must match the transportation question. A policy can support prediction; identified utility parameters or a reward shown to remain stable across specified settings can support behavioral interpretation; and an assignment or equilibrium model is additionally required when travelers affect one another through congestion or crowding. We call a model \emph{behaviorally disciplined} when learned components are embedded within declared feasible opportunities, utility or reward normalization, observation assumptions, and any network feedback required by the intended claim. The proposed hybrid uses ML for scale, context, and data integration while retaining these distinctions.
\end{insightbox}

The formal bridge is clearest for route choice. Recursive logit, logit dynamic discrete choice, and entropy-regularized decision policies share a soft Bellman recursion under explicit assumptions. On an absorbing road network, this recursion yields a globally normalized path distribution and a reward-learning interpretation \citep{fosgerau2013link,mai2015nested,mai2020relation,ziebart2008maxent}. Daily activity schedules extend the state to purpose, timing, duration, location, mode, resources, and household commitments, making activity choice a larger time--space sequential problem with endogenous goals and coupled constraints \citep{bhat1999activity,bowman2001activity,zimmermann2018activity,song2024stateirl}.

\rev{The mathematical unification stops at behavioral interpretation. Structural choice models may target utility parameters, elasticities, welfare, and intervention response; IRL learns a generally non-unique reward; IL reproduces a policy or occupancy; and offline RL optimizes a supplied objective. Observation error, unavailable alternatives, selection, and congestion can otherwise be absorbed into a flexible reward. The scientific question must therefore determine whether the estimand is utility, reward, policy, occupancy, constraint, or a trajectory distribution} \citep{ng1999shaping,train2009discrete}.

The review contributes a comparison linking random-utility and dynamic-choice models, IRL, IL, offline RL, and generative trajectory models to their estimands, assumptions, and defensible claims. It organizes evidence by learned object and strongest supported claim, proposes a hybrid architecture that preserves feasibility and behavioral meaning, and derives testable requirements for prediction, generalization, intervention, and system analysis. Sections~2--4 establish the scope, foundations, and formal framework; Sections~5--8 synthesize methods, evidence, architecture, and evaluation; the final sections state the implications, research agenda, limitations, and conclusions.

\section{Scope and Review Strategy}

This is a critical integrative review, not a systematic review with claimed exhaustive coverage. The focus is individual or household behavior represented as a sequence on a physical, temporal, or activity network. The core domains are road and transit route choice, pedestrian and bicycle routing, multimodal journeys, daily activity scheduling, and related dynamic discrete choices. \rev{Traffic control, autonomous driving, and vehicle routing are discussed only when their learning methods directly inform behavioral models estimated from observed trajectories.} The review also draws selectively on structural econometrics, operations research, and modern generative modeling.

Existing reviews cover important parts of this territory but ask different questions. Route-choice reviews emphasize choice-set generation and substitution, while recursive-model tutorials connect path choice with dynamic programming. Activity-travel reviews catalogue predictive ML; general IRL surveys organize inverse-learning algorithms; and transportation RL reviews focus mainly on prescriptive control and operations \citep{prato2009route,zimmermann2020tutorial,koushik2020activityml,arora2021survey,li2023rltransport,lai2024rltransport}. Table~\ref{tab:priorreviews} makes the incremental scope explicit. The present review does not compete with those catalogues on coverage. It focuses on the narrower cross-field question of which learned object supports which behavioral and transport-system claim.

\begin{longtable}{@{}P{0.20\textwidth}P{0.24\textwidth}P{0.24\textwidth}P{0.22\textwidth}@{}}
\caption{Positioning relative to adjacent reviews and tutorials.}\label{tab:priorreviews}\\
\toprule
\textbf{Review stream} & \textbf{Primary scope} & \textbf{Main contribution} & \textbf{Boundary addressed here} \\
\midrule
\endfirsthead
\toprule
\textbf{Review stream} & \textbf{Primary scope} & \textbf{Main contribution} & \textbf{Boundary addressed here} \\
\midrule
\endhead
\endfoot
\bottomrule
\endlastfoot
\citet{prato2009route} & \tabitem{Path-based random utility}\tabitem{Choice-set generation and route overlap} & \tabitem{Explains how sampled alternatives affect estimation}\tabitem{Compares overlap corrections and stochastic route-choice models} & \tabitem{Does not compare utility, reward, policy, and occupancy estimands}\tabitem{Does not extend the synthesis to activity schedules or ML-based trajectory learning} \\
\mobilityrowrule
\citet{zimmermann2020tutorial} & \tabitem{Recursive link/path choice}\tabitem{Inverse optimization on networks} & \tabitem{Derives the dynamic-programming likelihood}\tabitem{Connects recursive route models to inverse optimization} & \tabitem{Limited treatment of daily activity schedules and heterogeneous data sources}\tabitem{Does not center congestion feedback or intervention evidence} \\
\mobilityrowrule
\citet{koushik2020activityml} & \tabitem{Predictive ML for activity--travel behavior}\tabitem{Empirical application taxonomy} & \tabitem{Reviews prediction, interpretability, and transfer}\tabitem{Identifies practical activity-modeling applications} & \tabitem{Does not organize methods by learned behavioral object}\tabitem{Limited separation of preference, feasibility, observation error, and operator objectives} \\
\mobilityrowrule
\citet{arora2021survey} & \tabitem{General IRL theory and algorithms}\tabitem{Applications across domains} & \tabitem{Systematizes reward-learning assumptions and method families}\tabitem{Highlights ambiguity and evaluation issues} & \tabitem{No transport-specific treatment of choice sets, path overlap, or network equilibrium}\tabitem{Does not connect learned rewards to transport welfare quantities} \\
\mobilityrowrule
\citet{li2023rltransport,lai2024rltransport} & \tabitem{RL for transport operations and control}\tabitem{Signals, fleets, routing, and infrastructure} & \tabitem{Catalogues prescriptive objectives, algorithms, and deployment settings}\tabitem{Summarizes operational performance evidence} & \tabitem{Traveler reward is normally supplied rather than inferred}\tabitem{Observed behavior, preference recovery, and welfare interpretation are not the primary targets} \\
\mobilityrowrule
This review & \tabitem{Route and activity sequences learned from observed behavior}\tabitem{Choice, IRL, IL, offline RL, and generative methods} & \tabitem{Connects soft-Bellman, occupancy-flow, and inverse-problem formulations}\tabitem{Grades evidence by estimand and supported claim} & \tabitem{Adds observation, constraint, transfer, and equilibrium boundaries}\tabitem{\rev{Proposes a hybrid architecture and intervention-oriented evaluation framework}} \\
\end{longtable}

The search was updated through 15 August 2026 using Crossref, \rev{the Transportation Research International Documentation (TRID) database}, publisher search interfaces, targeted scholarly-web queries, and citation tracing. Queries combined transportation terms (route, activity, schedule, mobility), sequential-model terms (recursive, dynamic, Markov, trajectory, occupancy), and learning terms (IRL, imitation, offline, transfer, graph, transformer, diffusion, language model). Screening checked title/abstract relevance, confirmed the learned object and transportation task, and coded the data regime, evaluation design, transfer or intervention test, and publication status. Foundational papers establish formal bridges; prediction-only and frontier work is retained only when it clarifies a transport bottleneck or interpretive boundary \citep{prato2009route,arora2021survey,hussein2017imitation,levine2020offline}.

Included work either estimates utility, reward, policy, occupancy, constraints, or a trajectory distribution from behavior, or establishes a directly relevant identification, computational, or evaluation result. Claims are graded conservatively. Held-out trajectories estimate predictive performance for the evaluated population, network, and actions represented in the data; domain shifts test generalization or transfer; and interventions or data with known generating parameters are needed to assess recovery of utility, reward, or constraints. \emph{Established transport evidence} was evaluated on transport data or a transport network; a \emph{transferable method} was demonstrated elsewhere under assumptions that map to transportation; an \emph{emerging proposal} lacks substantial transport-specific validation. The procedure is documented at a conceptual level but is not exhaustive, independently duplicated, PRISMA-based, or reproducible as a systematic-review protocol.

\section{Sequential Mobility Choice: Transportation Foundations}

\rev{This section establishes transportation models that provide relevant baselines for many learning-based alternatives.} It proceeds from complete-path random utility to recursive link choice, flow-based choice, and activity-schedule models. These formulations do more than provide prediction baselines: they specify feasible alternatives, behavioral quantities, substitution patterns, and policy counterfactuals. At the same time, each has a characteristic bottleneck--path enumeration, repeated dynamic programs, aggregation, or combinatorial state. Accordingly, each subsection uses the same comparison: the model's behavioral advantage, its statistical or computational limitation, and the role in which ML can add value. The relevant question is not whether a neural model fits better in the abstract, but whether learned components expand representation, scale, or data coverage while retaining the choice set, normalization, constraints, and counterfactual meaning that the application requires.

\subsection{Path-based random utility}

Let $\mathcal{P}_{od}$ denote feasible paths between origin $o$ and destination $d$. A path-based random-utility model writes
\begin{equation}
 U_{in}=V_{in}+\varepsilon_{in}, \qquad
 V_{in}=\boldsymbol{\beta}^{\top}\mathbf{x}_{in},
 \label{eq:pathrum}
\end{equation}
where $\mathbf{x}_{in}$ contains travel time, cost, turns, road class, reliability, or other attributes. Under independent type-I extreme-value errors, the probability of choosing path $i$ from choice set $C_n$ is multinomial logit. \rev{With a correctly specified choice set, utility function, scale normalization, and treatment of endogeneity, this formulation can support interpretable marginal utilities, elasticities, values of time, and welfare analysis} \citep{mcfadden1974conditional,benakiva1985discrete,train2009discrete}.

The difficulty is the alternative space. Real networks contain enormous numbers of feasible, overlapping paths, and the analyst does not observe which paths a traveler considered. Generated choice sets can produce inconsistent inference unless their sampling mechanism is handled correctly \citep{frejinger2009sampling}. Shared links also induce correlation across path utilities, motivating path-size corrections, subnetworks, generalized extreme-value models, mixed logit, and other structures \citep{frejinger2007subnetworks,prato2009route}. Static path utilities are transparent, but complete-path enumeration and time-varying adaptation remain awkward.

Relative to a black-box path or sequence predictor, path-based random utility offers a transparent likelihood, interpretable marginal utilities, explicit substitution, and direct welfare calculations. Its limitations are a hand-specified utility surface and a sampled or generated alternative set whose omissions can bias estimation. ML can enrich path attributes, learn nonlinear residual utility and traveler heterogeneity, improve map matching, or propose informative alternatives; hybrid discrete-choice models demonstrate how learned representations or taste functions can be embedded inside a normalized utility model \citep{sifringer2020representation,han2022tastenet}. Those components should enter a sampling-corrected choice likelihood; otherwise improved prediction may come at the cost of unknown choice-set support, unnormalized scores, and uninterpretable elasticities \citep{frejinger2009sampling,train2009discrete}.

\subsection{Recursive route choice}

Recursive models replace one path choice with a sequence of link choices. At state $s$--a node or current link--the traveler selects a feasible outgoing action $a$. \rev{The action yields instantaneous utility $v_\theta(s,a)$ and transitions the traveler to another network state; the process continues until the absorbing destination is reached.} For i.i.d. extreme-value shocks with scale $\mu$, expected maximum utility satisfies a log-sum recursion. The induced local probabilities define a distribution over all feasible paths without enumerating them \citep{fosgerau2013link}.

Nested recursive logit allows state- or link-specific scales and richer substitution patterns \citep{mai2015nested}. \rev{Generalized recursive models embed multivariate extreme-value (MEV) correlation structures in the value recursion, and dynamic-programming algorithms make large network-based MEV specifications estimable} \citep{mai2016correlation,mai2017mev}. Decomposition and linear-algebra methods reduce repeated solution cost during estimation, while convergence analysis clarifies when undiscounted cyclic networks admit a well-defined value function \citep{mai2018decomposition,mai2022undiscounted}. Recursive models also extend to stochastic, time-dependent networks in which the traveler chooses a routing policy conditional on evolving information rather than committing to one deterministic path \citep{mai2021stochasticroute}. These developments matter for IRL because correlated shocks, changing transition conditions, and adaptive policies alter what a recovered reward can explain. \rev{Recursive route-choice models can be represented as parametric MDPs estimated from observed route decisions. Under the corresponding random-utility and transition assumptions, they can also be interpreted as structured entropy-regularized inverse-control models on a known network graph.}

This formulation solves path enumeration, but not every behavioral problem. The Markov state must contain all information required for continuation decisions. Omitted memory, habitual plans, unobserved information, and route-level attributes can violate that condition. Linkwise independent shocks also impose substitution restrictions, while cyclic networks require attention to improper looping. These are modeling assumptions, not implementation details \citep{puterman1994mdp,prato2009route,mai2022undiscounted}.

Correlation has several noninterchangeable meanings in sequential mobility choice. \rev{Shared links create correlation among complete-path utilities; persistent traveler effects create serial correlation across repeated trips; and nested or generalized extreme-value shocks alter local substitution} \citep{frejinger2007subnetworks,prato2009route,train2009discrete,mai2015nested,mai2016correlation,mai2017mev}. \rev{Congestion or coordination instead creates dependence across decision makers} \citep{ameli2022departure,shou2022markovrouting}. A neural embedding can improve prediction while leaving these sources unidentified. A behavioral application should state which correlation is represented, whether it enters utility, scale, latent classes, transitions, or equilibrium, and what substitution test distinguishes the specification from an independence model.

Relative to unconstrained next-link prediction, recursive choice enforces the declared feasible action set, produces destination-conditioned full-path probabilities, and provides a value recursion that can support behavioral parameters and network counterfactuals under the stated specification. Its main limitations are state sufficiency, restrictive utility or shock specifications, and the repeated solution of destination-specific dynamic programs at scale. Graph encoders and neural residual rewards can represent spatial context and heterogeneity, while learned value approximations reused across destinations, graph compression, and warm-started solvers can reduce computation \citep{zhao2023deepirl,barnes2023maps,battaglia2018relational}. The action mask, terminal condition, probability normalization, and exact or certified Bellman layer should remain explicit; a locally accurate neural policy need not reach the destination or preserve substitution under a closure \citep{fosgerau2013link,mai2022undiscounted}.

\subsection{Flow-based route choice}
Perturbed utility route choice (PURC) changes the unit of analysis from a selected path to a feasible amount of flow on every link. For one origin--destination pair, a stylized formulation is
\begin{equation}
 \max_{\mathbf{x}\in\mathcal{F}_{od}}
 \left\{\mathbf{u}^{\top}\mathbf{x}-\Omega(\mathbf{x})\right\},
 \label{eq:purc}
\end{equation}
where $\mathbf{x}$ contains nonnegative link flows, $\mathcal{F}_{od}$ imposes origin supply, destination demand, and conservation at intermediate nodes, $\mathbf{u}$ contains systematic link utilities, and the convex function $\Omega$ discourages all demand from collapsing onto one shortest or highest-utility route. Solving Equation~\eqref{eq:purc} therefore allocates demand directly over the network. No path set is generated, and routes that share links are coupled through the same link-flow variables. These features give PURC an overlap-sensitive, full-network representation and allow estimation or prediction to exploit convex optimization \citep{fosgerau2022purc}. Extensions can load many origin--destination demands and update travel times, connecting regularized individual choice to stochastic traffic assignment and congestion feedback \citep{yao2024pusta}.

The comparison with path-based and recursive models is consequential. A path-based logit model assigns probabilities to enumerated complete routes and offers direct path-level likelihoods, but requires a sampled or generated choice set. Recursive choice avoids path enumeration by assigning probabilities to successive links and retains an explicit individual policy, but repeatedly solves a destination-conditioned value function. PURC instead solves for conserved link flows in one optimization problem. This is especially attractive when the scientific target is aggregate network loading, scalable assignment, or substitution among overlapping routes. Its limitation is individual-level resolution: the same aggregate link-flow vector can be decomposed into paths in multiple ways, so it need not identify a unique path distribution, traveler-specific taste heterogeneity, or the reward process that generated the flow. The perturbation, demand units, and interpretation of fractional flow must therefore be declared rather than treated as a generic stochastic-choice error.

The flow view also creates a mathematical bridge to occupancy-based imitation learning because expected state--action visitation obeys an analogous conservation equation. That bridge is easiest for route choice, where each action moves flow through a physical network toward an absorbing destination. It is harder to transfer unchanged to activity choice. An activity flow must live on a much larger time--space--resource graph, distinguish starting, continuing, and ending an episode, account for duration, and preserve mandatory activities, household coordination, and end-of-day completion. Aggregating these states can erase the very timing and resource constraints that make an activity schedule behaviorally meaningful; retaining them can make the flow program extremely large. \rev{Activity applications therefore need semi-Markov occupancies or time-expanded states that record clock time, location, episode duration, and remaining commitments; structured decompositions may be needed to keep this representation computationally manageable} \citep{puterman1994mdp,zimmermann2018activity}.

ML can learn nonlinear edge utilities, demand segments, or differentiable equilibrium surrogates, while occupancy-ratio methods can reweight logged flows across regimes \citep{nachum2019dualdice,yao2024pusta}. However, flow conservation, schedule/resource feasibility, and congestion feedback should remain explicit solver constraints. A low-error neural flow forecast can otherwise violate network balance or reproduce aggregate counts while obscuring the behavioral mechanism needed for a policy counterfactual.

\subsection{Activity-based and schedule choice}

Activity-based demand models replace independent trips with the organization of a day. Classical systems combine activity-pattern, tour, destination, mode, and time-of-day models; later microsimulators use rules, hazards, optimization, or network representations to construct schedules \citep{bhat1999activity,bowman2001activity,miller2003tasha}. Their behavioral advantage is substantive: trips are linked because they serve activities, household members coordinate, and time spent in one episode changes later opportunities.

\rev{Activity scheduling can be represented as sequential learning on a suitably specified time--space network.} A state may be written
\begin{equation}
 s_t=(\ell_t,t,\mathbf{b}_t,\mathbf{h}_t,\mathbf{c}_n),
 \label{eq:activitystate}
\end{equation}
where $\ell_t$ is location, $\mathbf{b}_t$ records remaining time, money, or energy budgets, $\mathbf{h}_t$ summarizes completed or required activities, and $\mathbf{c}_n$ contains personal and household context. Actions include staying, traveling, selecting a mode or destination, starting an activity, and ending the day. A complete schedule becomes a path through this augmented network. Mixed recursive logit has been estimated on such networks for joint activity, location, timing, and mode choice \citep{zimmermann2018activity}; recent IRL applications learn activity-travel rewards and policies from mobile traces or travel surveys \citep{song2024stateirl,liang2026interpretable}.

Activity choice also exposes the limits of a naive MDP. An activity episode has a type, location, start time, and duration, and its continuation changes the opportunity set for the rest of the day. A semi-Markov representation makes duration explicit: choosing activity $a_t$ and duration $\Delta_t$ moves clock time from $t$ to $t+\Delta_t$ and accumulates activity benefit, schedule delay, and travel cost over that interval. Without such a representation, a model may reproduce the next activity while misrepresenting time allocation. Long-term commitments, household negotiation, and flexible timing may further require partially observed beliefs or multi-agent models \citep{puterman1994mdp,zimmermann2018activity}.

The empirical base is expanding but remains thinner than for route choice. Data-driven activity schedulers have generated regional populations with machine-learning components while retaining activity-based validation targets \citep{drchal2019scheduler}. State-based IRL has been evaluated on cellular signaling data with reward recovery and transfer tasks, and context-aware IRL has generated daily schedules from a large resident travel survey \citep{song2024stateirl,liu2025contextirl}. Interpretable deep IRL has also used policy distillation and reward analysis on Singapore travel-survey sequences \citep{liang2026interpretable}. These studies demonstrate prediction and representation, but they do not yet establish that learned rewards support welfare analysis or remain invariant under major land-use, household, or accessibility changes.

Activity-schedule models preserve the derived nature of travel, whole-day feasibility, time and resource accounting, and linkages among activity, location, mode, and duration choices. Their limitation is the combinatorial state and alternative space, compounded by sparse diaries, partial observation, long memory, heterogeneous constraints, and household interaction. Representation learning can compress histories and context; latent-variable and data-fusion models can combine surveys with passive traces; and sequence or IRL methods can learn flexible schedule rewards and policies \citep{drchal2019scheduler,song2024stateirl,liang2026interpretable}. \rev{Claims about complete activity schedules remain limited unless duration, resource transitions, mandatory episodes, household coupling where relevant, and end-of-day completion are represented and evaluated explicitly} \citep{puterman1994mdp,zimmermann2018activity}. Next-activity accuracy alone is not evidence of a feasible daily schedule or a transferable behavioral reward.

\begin{longtable}{@{}P{0.18\textwidth}P{0.34\textwidth}P{0.38\textwidth}@{}}
\caption{Route and activity choice as distinct instances of sequential mobility choice.}\label{tab:instances}\\
\toprule
\textbf{Dimension} & \textbf{Route-choice instance} & \textbf{Activity-choice instance} \\
\midrule
\endfirsthead
\toprule
\textbf{Dimension} & \textbf{Route-choice instance} & \textbf{Activity-choice instance} \\
\midrule
\endhead
\endfoot
\bottomrule
\endlastfoot
State & \tabitem{Current node/link and destination}\tabitem{Departure time, information, vehicle/mode, and traveler context} & \tabitem{Location, clock time, and current activity}\tabitem{Past episodes, remaining commitments/resources, and household context} \\
\mobilityrowrule
Action & \tabitem{Choose a feasible outgoing link, turn, transit service, or maneuver}\tabitem{Possibly revise the route as information changes} & \tabitem{Travel, wait, begin, continue, or end an activity}\tabitem{Choose location, mode, duration, and coordination action} \\
\mobilityrowrule
Horizon & \tabitem{Usually one origin--destination trip}\tabitem{Ends at an absorbing destination; minutes to hours} & \tabitem{Day or week with multiple episodes}\tabitem{Endogenous goals, explicit durations, and several terminal obligations} \\
\mobilityrowrule
Dependence & \tabitem{Shared-link overlap and route persistence}\tabitem{Information updates, congestion, and unobserved taste heterogeneity} & \tabitem{Duration dependence and trip chaining}\tabitem{Accessibility, household coupling, resource budgets, and congestion} \\
\mobilityrowrule
Primary observations & \tabitem{GPS or probe trajectories and map matches}\tabitem{Stated routes, link counts, travel times, and incidents} & \tabitem{Travel/activity diaries and time-use surveys}\tabitem{Smart cards, phone traces, household attributes, and population margins} \\
\mobilityrowrule
Core validation & \tabitem{Complete-path likelihood and destination completion}\tabitem{Overlap/substitution, calibration, and response to tolls or closures} & \tabitem{Complete-day feasibility and terminal obligations}\tabitem{Episode timing/duration, joint household schedules, and accessibility-policy response} \\
\end{longtable}

\section{\rev{A Unified Sequential Mobility Choice Framework}}

\rev{This section provides the formal basis of the review through three distinct connections. The soft-Bellman bridge relates recursive choice to entropy-regularized decision policies. The occupancy-flow bridge relates trajectory visitation to network flow conservation. The inverse-problem bridge relates observed choices to latent utility or reward without assuming that the resulting quantities have identical behavioral meanings.} The construction begins with a four-layer generative model so that environment dynamics, behavioral objectives, stochastic choice, and data recording remain distinct. It then states the assumptions under which the bridges hold and the identification boundaries beyond which predictive equivalence does not support a behavioral interpretation. This organization draws on standard MDP and dynamic-choice formulations, recursive route choice, maximum-entropy inverse decision making, and occupancy-based imitation \citep{puterman1994mdp,rust1987replacement,fosgerau2013link,ziebart2008maxent,ziebart2010causal,mai2020relation,ho2016gail}.

\subsection{Four layers}

Let a context $c$ contain the destination, traveler attributes, trip purpose, time period, information regime, and policy environment. A latent trajectory is $\tau=(s_0,a_0,s_1,\ldots,s_T)$. Its model probability can be factored as
\begin{equation}
 p(y,\tau\mid c)=p(s_0\mid c)
 \prod_{t=0}^{T-1}\pi_\theta(a_t\mid s_t,c)
 P(s_{t+1}\mid s_t,a_t,c)\,p_\psi(y\mid\tau,c),
 \label{eq:fourlayer}
\end{equation}
where $y$ is the recorded GPS, diary, smart-card, or cellular observation. \rev{The behavioral objective and stochastic choice mechanism jointly induce the policy $\pi_\theta$. Reward therefore affects Equation~\eqref{eq:fourlayer} through the policy rather than through a separate probability factor.} Equation~\eqref{eq:fourlayer} combines the standard controlled Markov factorization with an explicit observation model, thereby separating latent behavior from the process that records it \citep{puterman1994mdp,ziebart2010causal,mai2023incomplete}.

\begin{longtable}{@{}P{0.17\textwidth}P{0.25\textwidth}P{0.25\textwidth}P{0.23\textwidth}@{}}
\caption{Four layers of sequential mobility choice.}\label{tab:fourlayers}\\
\toprule
\textbf{Layer} & \textbf{Transportation specification} & \textbf{ML representation} & \textbf{Operational role and diagnostic} \\
\midrule
\endfirsthead
\toprule
\textbf{Layer} & \textbf{Transportation specification} & \textbf{ML representation} & \textbf{Operational role and diagnostic} \\
\midrule
\endhead
\endfoot
\bottomrule
\endlastfoot
Environment & \tabitem{Network connectivity, schedules, operating rules, and available modes}\tabitem{Time, congestion, resources, and terminal conditions} & \tabitem{State/action spaces and transition kernel}\tabitem{Action masks, horizon, and simulator} & \tabitem{Defines what can physically occur}\tabitem{Diagnose invalid moves, unreachable destinations, and violated time/resource transitions} \\
\mobilityrowrule
Objective & \tabitem{Travel-time and monetary disutility}\tabitem{Activity benefit, reliability, comfort, risk, and schedule delay} & \tabitem{Reward/return and cost functions}\tabitem{Constraint costs, budgets, and learned residuals} & \tabitem{Ranks feasible sequences and defines trade-offs}\tabitem{Diagnose scale, shaping, sign, and reward--constraint confounding} \\
\mobilityrowrule
Choice mechanism & \tabitem{Random-utility shocks, scale, nesting, and taste heterogeneity}\tabitem{Information and expectation formation} & \tabitem{Policy, planner, and value function}\tabitem{Entropy regularization and probability normalization} & \tabitem{Converts values into choice probabilities; occupancies are derived after combining the policy with transitions}\tabitem{Diagnose calibration, substitution, path overlap, and unsupported extrapolation} \\
\mobilityrowrule
Observation & \tabitem{GPS, diaries, smart cards, counts, and survey weights}\tabitem{Map matching, missing trips, sampling, and selection} & \tabitem{Demonstrations and latent trajectories}\tabitem{Sensor, missingness, and behavior-policy models} & \tabitem{Maps latent behavior to recorded data}\tabitem{Diagnose selection bias, ambiguous paths, measurement error, and source disagreement} \\
\end{longtable}

This decomposition prevents common category errors. Road connectivity belongs to the environment, not the reward. A legal prohibition is not necessarily a large negative preference. GPS error is not behavioral randomness. Entropy can represent unobserved utility variation, bounded precision, or a deliberate regularizer, but those interpretations imply different counterfactuals. Heterogeneity may enter rewards, beliefs, constraints, or noise scale; placing it indiscriminately in a neural embedding weakens behavioral meaning \citep{benakiva1985discrete,train2009discrete,altman1999cmdp,ziebart2010causal}.

\begin{figure}[t]
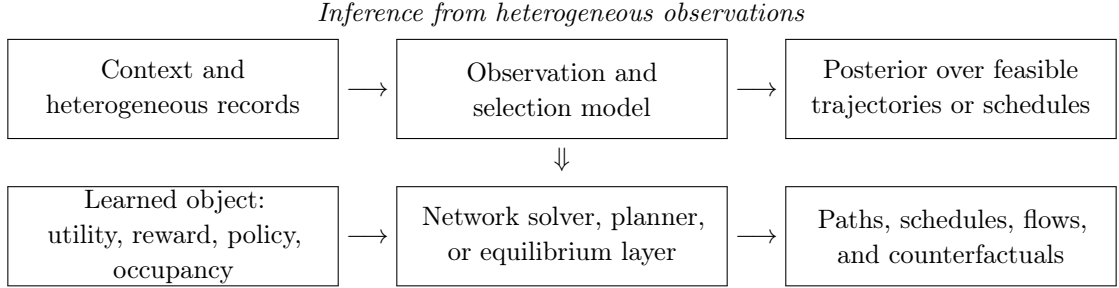

\centering
\small
\begin{tabular}{c@{\;$\longrightarrow$\;}c@{\;$\longrightarrow$\;}c}
\multicolumn{3}{c}{\rev{\emph{Inference from heterogeneous observations}}}\\[0.25em]
\fbox{\parbox[c][3.1em][c]{0.24\textwidth}{\centering \rev{Context and\\heterogeneous records}}} &
\fbox{\parbox[c][3.1em][c]{0.24\textwidth}{\centering \rev{Observation and\\selection model}}} &
\fbox{\parbox[c][3.1em][c]{0.24\textwidth}{\centering \rev{Posterior over feasible\\trajectories or schedules}}} \\
\multicolumn{3}{c}{$\Downarrow$} \\[0.35em]
\fbox{\parbox[c][3.1em][c]{0.24\textwidth}{\centering Learned object:\\utility, reward, policy, occupancy}} &
\fbox{\parbox[c][3.1em][c]{0.24\textwidth}{\centering Network solver, planner,\\or equilibrium layer}} &
\fbox{\parbox[c][3.1em][c]{0.24\textwidth}{\centering Paths, schedules, flows,\\and counterfactuals}}
\end{tabular}
\caption{\rev{The sequential-mobility-choice inference and decision pipeline. The top arrows show inference from heterogeneous records to a posterior over latent feasible behavior; the corresponding generative observation model runs in the reverse direction. The scientific objective determines what is learned and which solver converts it into individual and system outcomes.}}
\label{fig:pipeline}
\end{figure}

\subsection{The soft Bellman bridge}

For reward $r_\theta(s,a,c)$, transition kernel $P$, discount factor $\gamma$, and entropy scale $\mu>0$, define
\begin{align}
 Q_\theta(s,a,c) &= r_\theta(s,a,c)+\gamma
 \mathbb{E}_{s'\sim P(\cdot\mid s,a,c)}[V_\theta(s',c)], \label{eq:softq}\\
 V_\theta(s,c) &= \mu\log\sum_{a\in\mathcal{A}(s,c)}
 \exp\{Q_\theta(s,a,c)/\mu\}, \label{eq:softv}\\
 \pi_\theta(a\mid s,c) &=
 \exp\{[Q_\theta(s,a,c)-V_\theta(s,c)]/\mu\}. \label{eq:softpolicy}
\end{align}
In dynamic logit, Equation~\eqref{eq:softv} is the expected maximum utility generated by extreme-value shocks. In entropy-regularized RL it is a soft value function. In maximum-causal-entropy IRL, the likelihood of demonstrations is induced by Equation~\eqref{eq:softpolicy}. More precisely, entropy-regularized MDPs and stochastic MDPs with independent Gumbel reward shocks--the logit dynamic-discrete-choice formulation--are equivalent in both forward-control and inverse-learning perspectives \citep{mai2020relation}. Recursive logit is the network specialization in which actions are outgoing links, transitions are usually deterministic, the destination is absorbing, and the horizon is an undiscounted stochastic-shortest-path problem \citep{rust1987replacement,fosgerau2013link,ziebart2010causal,zimmermann2020tutorial}.

For the undiscounted episodic case ($\gamma=1$), consider a feasible path $\sigma=(s_0,a_0,\ldots,s_T=d)$ with deterministic transitions, additive reward, and $V(d)=0$. The product of local policies telescopes:
\begin{equation}
 p_\theta(\sigma\mid s_0,d)=
 \exp\left\{\frac{1}{\mu}\sum_{t=0}^{T-1}r_\theta(s_t,a_t,d)
 -\frac{1}{\mu}V_\theta(s_0,d)\right\}.
 \label{eq:telescoping}
\end{equation}
Thus local soft choices induce a globally normalized exponential distribution over feasible paths. \rev{Under the assumptions stated above, this establishes a formal connection among recursive logit, logit DDC, and maximum-entropy route IRL. Recursive logit supplies the network state--action structure, while the entropy-regularized/stochastic-MDP equivalence supplies the common soft policy and its forward and inverse interpretations} \citep{mai2020relation}. It requires a proper stochastic-shortest-path model: the destination must be reached with probability one and rewards must rule out infinitely attractive cycles so that the partition function is finite. With discounting, stochastic transitions, finite-horizon time indexing, nested scales, non-Markovian attributes, or alternative entropy regularizers, the corresponding trajectory distribution changes and the displayed telescoping form does not apply unchanged. Maximum trajectory entropy and maximum causal entropy must likewise be distinguished when transitions are stochastic \citep{ziebart2008maxent,ziebart2010causal,mai2022undiscounted,zimmermann2020tutorial}.

\subsection{The occupancy-flow bridge}

For initial distribution $\rho_0$ and policy $\pi$, define the discounted state-action occupancy
\begin{equation}
 d_\pi(s,a)=(1-\gamma)\sum_{t=0}^{\infty}\gamma^t
 \Pr_\pi(s_t=s,a_t=a).
\end{equation}
It satisfies
\begin{equation}
 \sum_a d_\pi(s,a)=(1-\gamma)\rho_0(s)
 +\gamma\sum_{s',a'}P(s\mid s',a')d_\pi(s',a').
 \label{eq:occupancy}
\end{equation}
The quantity $d_\pi(s,a)$ is the probability mass assigned to taking action $a$ while visiting state $s$ when trajectories are generated by policy $\pi$, with later visits geometrically downweighted by $\gamma$. \rev{Summing it over actions gives state visitation, while summing it over states gives the marginal discounted frequency of an action label when that label is shared across states.} Equation~\eqref{eq:occupancy} says that mass at a state equals externally supplied initial mass plus discounted mass arriving from predecessor state--action pairs. It is therefore a flow-conservation equation with source terms. In route choice, occupancies yield expected link or turn usage and, after multiplication by demand, predicted traffic flow. \rev{In activity choice, they summarize visitation frequencies over time--location--activity states, supporting comparisons of schedule patterns and time spent in different opportunity contexts.} In IL, matching expert and learner occupancies reproduces the frequency of behavior over the whole trajectory rather than only the next action; adversarial IL implements this principle through distribution matching \citep{puterman1994mdp,ho2016gail}. PURC, entropy-regularized assignment, and occupancy matching consequently operate on closely related feasible polytopes \citep{fosgerau2022purc}. Their objectives still differ: one may maximize perturbed utility, another minimize a divergence from expert occupancy, and another recover rewards whose optimal occupancy matches data.

The analogy becomes exact only after matching units. For a finite trip, define the unnormalized episodic visitation measure
\begin{equation}
 \bar d_\pi(s,a\mid c)=
 \mathbb{E}_\pi\!\left[\sum_{t=0}^{T-1}
 \mathbf{1}\{s_t=s,a_t=a\}\mid c\right].
 \label{eq:episodicoccupancy}
\end{equation}
If $q_c$ travelers share context $c$, then $q_c\bar d_\pi(s,a\mid c)$ is an expected link-turn flow. The normalized discounted measure in Equation~\eqref{eq:occupancy} is instead a probability distribution useful for continuing tasks and density-ratio estimation. For activity models, decision-epoch occupancy counts visits; it represents time spent only after weighting by episode duration or using a semi-Markov occupancy. These distinctions prevent probability mass, person trips, vehicles, and person-time from being conflated \citep{puterman1994mdp,zimmermann2018activity,nachum2019dualdice}.

\subsection{Inverse learning and the limits of behavioral identification}

For linear reward $r_\theta(s,a)=\theta^{\top}f(s,a)$, the maximum-entropy log-likelihood gradient has the familiar form
\begin{equation}
 \nabla_\theta\ell(\theta)=
 \widehat{\mathbb{E}}_{E}\!\left[\sum_t f(s_t,a_t)\right]
 -\mathbb{E}_{\pi_\theta}\!\left[\sum_t f(s_t,a_t)\right],
 \label{eq:featurematching}
\end{equation}
which matches demonstrated and model feature counts \citep{ziebart2008maxent}. \rev{Structural DDC likewise infers per-period utility from conditional choice probabilities and continuation values} \citep{hotz1993ccp,aguirregabiria2010dynamic}. Both are inverse problems: choices reveal payoffs only relative to dynamics, opportunity sets, information, and shock assumptions.

The boundary is identification. Rewards are not generally unique. \rev{Potential-based shaping can preserve optimal policies; reward rescaling can be observationally equivalent when paired with a corresponding noise-scale normalization; and constant shifts are innocuous only under fixed-horizon or otherwise controlled episode-length conditions} \citep{ng1999shaping,ng2000algorithms}. Omitted states can cause a learned reward to absorb beliefs, constraints, selection, or equilibrium conditions. In route choice, time and cost coefficients support a value of time only after scale normalization and under exogeneity \citep{train2009discrete}. In activity choice, an inferred preference for staying home may instead reflect childcare, inaccessible modes, or missing activities. IRL does not remove these structural problems; it makes their sequential form visible \citep{malik2021icrl}.

\begin{insightbox}
\textbf{Identification remark.} Formal equivalence does not imply inferential equivalence. Two methods may generate the same policy while assigning different meanings to reward, shocks, regularization, and unobserved constraints. Behavioral claims require normalizations and invariance assumptions beyond predictive fit.
\end{insightbox}

\subsection{Running example: a constrained commute and activity chain}
\label{sec:runningexample}

Consider a stylized weekday decision that will serve as a running example. A traveler begins at home, must drop a child at school before a fixed deadline, must reach work within a preferred arrival window, may stop for groceries after work, and then returns home. The realized chain is therefore not just a route but a linked sequence such as home $\rightarrow$ school $\rightarrow$ work $\rightarrow$ grocery $\rightarrow$ home. At each episode the traveler chooses whether to start or end an activity, when to depart, which mode to use, and which road or transit link to take. A policy experiment introduces a cordon toll and a temporary road closure; because many travelers respond, travel times and crowding change endogenously.

A unified state records location, time, current activity, remaining commitments, available modes and vehicles, accumulated expenditure, and relevant network information. Connectivity, operating hours, one-way rules, and unavailable modes define the feasible action set. \rev{A school deadline may be encoded as a hard schedule requirement, while daily time, fare or toll expenditure, and vehicle range can be tracked as resource states or represented through an appropriate constrained Markov decision process (CMDP).} \rev{Travel time, monetary cost, reliability, schedule delay, activity duration, safety, and comfort enter an interpretable reward specification}, while a graph or sequence representation may capture nonlinear neighborhood context, history, and heterogeneous perceptions. This separation matters: bypassing a tolled link could reveal price sensitivity, a school deadline, lack of vehicle access, incomplete information, a GPS gap, or congestion caused by other travelers. An unrestricted reward model can absorb all six explanations.

The observation record is also layered. A household survey may reveal trip purpose, the school commitment, available vehicles, and socioeconomic context but cover only one or two days. GPS provides repeated fine-grained routes with missing segments and weak semantics; a smart card reveals transit transactions; traffic counts and probes observe aggregate flow and speed; and the network supplies legal transitions and tolls. \rev{In Equation~\eqref{eq:fourlayer}, record-linked sources may inform the same latent activity--travel trajectory through separate observation and selection models. Unlinked samples and aggregate counts instead constrain a shared population distribution or its margins; they should not be concatenated as if they were equally reliable records of one traveler} \citep{mai2023incomplete,graellsgarrido2023fusion,vo2026fusion}.

\rev{The same observed commute supports several distinct questions, but not the same conclusion.}
\begin{insightbox}
\textbf{\rev{Three readings of the same detour.}}
\begin{itemize}
\item \rev{A recursive or dynamic choice model asks how time, toll, reliability, and schedule delay change the probability of each feasible decision. With valid availability, exogeneity, normalization, observation, and equilibrium assumptions, the estimated trade-offs may support elasticities, value of time, and welfare analysis.}
\item \rev{IL asks which activity, mode, or link comes next for states and actions represented in the data. IRL instead asks which reward makes the observed chain probable and can be used for replanning. Neither predictive accuracy nor reward recovery alone establishes economic utility or welfare.}
\item \rev{Offline RL asks which recommendation improves a supplied operator objective, while a multi-agent or equilibrium model asks how many interacting responses change congestion and subsequent incentives. These models require separate traveler-acceptance evidence and an empirically defensible system-response model.}
\end{itemize}
\end{insightbox}

\rev{Thus, a toll detour may reveal a conditional time--cost trade-off, reproduce a local policy, express a reward under stated invariance assumptions, or implement an operator objective. Explanation, prediction, transfer, simulation, and prescription require different estimands and tests} \citep{train2009discrete,arora2021survey,levine2020offline}.

\section{What Is Learned? A Map of Method Families}

Methods are classified by their learned object rather than their algorithmic label. \rev{GPS traces or diaries can be used to estimate utility coefficients, rewards, policies, occupancies, constraints, or trajectory distributions under method-specific identification assumptions.} Each object has different data requirements and permissible claims. \rev{Table~\ref{tab:methods} provides the paper's main comparison of these objects; the running commute above illustrates why their interpretations differ.}

\begin{longtable}{@{}P{0.17\textwidth}P{0.17\textwidth}P{0.22\textwidth}P{0.27\textwidth}@{}}
\caption{Method families classified by their primary estimand.}\label{tab:methods}\\
\toprule
\textbf{Family} & \textbf{Learned object} & \textbf{Best-supported use} & \textbf{Behavioral boundary / required check} \\
\midrule
\endfirsthead
\toprule
\textbf{Family} & \textbf{Learned object} & \textbf{Best-supported use} & \textbf{Behavioral boundary / required check} \\
\midrule
\endhead
\endfoot
\bottomrule
\endlastfoot
Path-based choice & \tabitem{Scale-normalized utility coefficients}\tabitem{Probability over a generated path set} & \tabitem{Interpretable time/cost trade-offs}\tabitem{Elasticities, substitution, and welfare on declared alternatives} & \tabitem{Correct for sampled choice sets and shared-link overlap}\tabitem{Test sensitivity to omitted feasible routes} \\
\mobilityrowrule
Recursive choice / DDC & \tabitem{\rev{Per-decision utility and shock scale}}\tabitem{Soft value and recursive policy} & \tabitem{\rev{Uses the paths induced by the declared network without explicit enumeration}}\tabitem{\rev{Supports dynamic counterfactuals under the model assumptions}} & \tabitem{State must contain relevant memory and availability}\tabitem{Repeated value solves and endogeneity remain} \\
\mobilityrowrule
PURC / flow choice & \tabitem{Perturbed utility/cost}\tabitem{Aggregate network flow} & \tabitem{Convex network-wide prediction}\tabitem{Direct connection to assignment and congestion models} & \tabitem{Individual histories and heterogeneity are aggregated}\tabitem{Person-level inference requires an additional population model} \\
\mobilityrowrule
Behavior cloning & \tabitem{Conditional next-action policy}\tabitem{Optional autoregressive sequence policy} & \tabitem{Fast prediction and large-scale simulation near observed states}\tabitem{No planner required at inference} & \tabitem{Local errors compound after leaving expert support}\tabitem{Action masks, uncertainty, and sequence-level validation are essential} \\
\mobilityrowrule
IRL & \tabitem{Reward or soft value representation}\tabitem{Policy induced by a planner} & \tabitem{Replanning for new destinations or environments}\tabitem{Potential recovery of stable behavioral trade-offs} & \tabitem{Fix scale and shaping conventions}\tabitem{Test reward invariance and separate observation, constraint, and congestion effects} \\
\mobilityrowrule
\rev{Constrained IRL / inverse constraint learning (ICRL)} & \tabitem{Reward under supplied constraints}\tabitem{Or latent constraint cost/budget under a supplied reward} & \tabitem{Separates preferences from feasibility and cumulative limits}\tabitem{Represents accessibility, safety, time, or resource restrictions} & \tabitem{Joint reward--constraint recovery is weakly identified}\tabitem{Requires regulations, stated availability, multiple tasks, or negative examples} \\
\mobilityrowrule
Occupancy IL & \tabitem{Policy through state-action occupancy matching}\tabitem{Sometimes an implicit discriminator reward} & \tabitem{Matches complete rollout visitation rather than one-step labels}\tabitem{Useful for realistic populations in a known scenario} & \tabitem{Matched current flows do not identify preferences}\tabitem{Transfer requires a credible simulator and invariance evidence} \\
\mobilityrowrule
\rev{Distribution Correction Estimation (DICE) / density-ratio learning} & \tabitem{Target-to-behavior occupancy ratio}\tabitem{Reweighted value, occupancy, or target policy} & \tabitem{Offline evaluation and imitation without new rollouts}\tabitem{Constrained optimization close to logged support} & \tabitem{Coverage must hold within context strata}\tabitem{Use episodic flow equations for finite trips; ratios are not tastes} \\
\mobilityrowrule
Observation-aware /\newline state-only learning & \tabitem{Posterior over latent paths or schedules}\tabitem{State or transition occupancy and implicit policy} & \tabitem{Uses disconnected GPS, coarse locations, and action-free records}\tabitem{Propagates map-match and missing-episode uncertainty} & \tabitem{Results depend on sensor, missingness, and selection models}\tabitem{Retain multiple paths when observations do not identify one} \\
\mobilityrowrule
Mixed-quality / avoidance learning & \tabitem{Quality-weighted reward, value, or policy}\tabitem{Avoidance rule from negative examples} & \tabitem{Combines trusted, suboptimal, unlabeled, and undesirable traces}\tabitem{Reduces the assumption that all observed behavior is expert} & \tabitem{Quality labels must correspond to a stated objective or constraint}\tabitem{Check sensitivity to mislabeled and selectively observed examples} \\
\mobilityrowrule
Offline RL & \tabitem{Policy maximizing a supplied operator return}\tabitem{Optional safety or resource constraints} & \tabitem{Prescriptive routing and recommendation from fixed logs}\tabitem{Can improve on undesirable historical behavior} & \tabitem{Unsupported actions create extrapolation error}\tabitem{Keep traveler preference, acceptance, and operator objective separate} \\
\mobilityrowrule
Sequence / generative model & \tabitem{Conditional distribution over paths or schedules}\tabitem{Long-context latent representation} & \tabitem{Multimodal generation, completion, and simulation}\tabitem{Captures history beyond a small Markov state} & \tabitem{Constrained decoding must enforce connectivity and time/resources}\tabitem{Plausibility does not establish utility or intervention validity} \\
\mobilityrowrule
LLM / generative agent & \tabitem{Semantic variables, prompted choice, persona, or high-level plan}\tabitem{Natural-language interface to tools} & \tabitem{Extracts context from diaries and policy text}\tabitem{Supports low-label prediction and exploratory simulation} & \tabitem{Prompt/model sensitivity, hallucination, privacy, and cost}\tabitem{\rev{Use typed outputs, feasibility-enforcing solvers, equal-information baselines, and versioned audits}} \\
\end{longtable}

ML contributes in three ways. It can reduce computation through reusable policies, shared representations, and stochastic optimization; support nonlinear specifications that share information across travelers while retaining individual differences; and make incomplete, state-only, or mixed-quality mobility records usable \citep{barnes2023maps,mai2023incomplete,pham2025iostom,hoang2024sprinql}. Hybrid choice models show how these components can enrich a normalized probabilistic core \citep{sifringer2020representation,han2022tastenet}.

These gains do not guarantee identification. Policies can fail outside support, occupancies can match flows without recovering preferences, and flexible rewards can absorb observation error, constraints, or congestion \citep{ross2011dagger,ho2016gail,ng1999shaping,levine2020offline}. \rev{Hereafter, we reserve \emph{structural} for quantities intended to retain a specified interpretation under a defined counterfactual; a graph or architectural constraint alone is insufficient. The named hybrid framework remains \emph{behaviorally disciplined}, as defined in the Introduction.}

\subsection{Direct policy learning}

Behavior cloning estimates $\pi(a\mid s,c)$ directly from labeled choices. It is appropriate for predicting the next link, mode, activity, or schedule event when deployment remains close to the observed states and actions. Its central weakness is distribution shift: one error can move the model to a state absent from training, causing subsequent predictions to deteriorate. Dataset Aggregation (DAgger) addresses this problem by asking an expert to label learner-visited states \citep{ross2011dagger}, but retrospective mobility datasets rarely permit such interaction. Offline applications therefore need action masks, uncertainty and data-coverage diagnostics, and destination- or plan-level losses in addition to local cross-entropy \citep{hussein2017imitation,arora2021survey}. Behavior cloning does not recover reward and should not be used for welfare analysis.

\subsection{Reward learning through IRL}

IRL specifies a behavioral model in which demonstrated actions become more probable as cumulative reward increases; the assumed degree and form of optimality vary across methods. Major approaches compare demonstrated and modeled feature counts, define normalized maximum-entropy trajectory distributions, or estimate rewards and values through adversarial or value-based objectives \citep{ng2000algorithms,abbeel2004apprenticeship,ratliff2006margin,ziebart2008maxent,fu2018airl,garg2021iqlearn}.

\rev{Reward learning is especially attractive when a specified reward component is hypothesized to remain invariant while destinations, networks, or constraints change. A road-type or safety reward can then be replanned for a new destination, and an activity reward can be evaluated in a changed transport system.} Recent route-choice work uses context-dependent deep rewards and demonstrates prediction to unseen destinations \citep{zhao2023deepirl}. Large-scale industrial work uses graph compression, spatial mixtures of experts, and receding-horizon inverse planning on road graphs with hundreds of millions of states \citep{barnes2023maps}.

The promise should be stated carefully. Transfer occurs only if the learned reward captures a stable preference rather than city-specific opportunity, platform selection, congestion, or sensor artifacts. Raw neural reward magnitudes are not economic utility coefficients. A responsible application reports which reward contrasts are identified, fixes scale and shaping conventions, tests invariance across environments, and separates reward interpretation from policy interpretation \citep{ng1999shaping,arora2021survey,train2009discrete}.

\subsection{Occupancy matching and adversarial imitation}

Generative adversarial imitation learning compares the state--action frequencies generated by complete expert and learner trajectories rather than only next actions \citep{ho2016gail}. For transportation, this aligns naturally with link flows, time--location profiles, and activity frequencies and can reduce errors that compound during rollout. However, occupancy matching can reproduce a current demand regime without recovering preferences that remain stable under new origin--destination matrices or capacities. It also requires a simulator validated for action feasibility and the network dynamics relevant to the application. The method is therefore most compelling for simulating a realistic population in a known scenario, rather than for inferring value of time or welfare \citep{hussein2017imitation,arora2021survey}.

\subsection{The DICE family: occupancy and density-ratio correction}

The distribution correction estimation (DICE) family makes the occupancy-flow bridge operational for fixed, off-policy datasets. Depending on the formulation, the target is a discounted episodic occupancy or a stationary distribution. Let $d_\beta(s,a)$ be the occupancy of the behavior policy that generated the observations and $d_\pi(s,a)$ the corresponding occupancy of a target policy. DICE methods estimate the ratio between target-policy and logged-data visitation frequencies,
\begin{equation}
 w_\pi(s,a)=\frac{d_\pi(s,a)}{d_\beta(s,a)}
 \label{eq:diceratio}
\end{equation}
by enforcing flow-conservation identities rather than multiplying a long sequence of action-level importance ratios. Under correct ratio estimation, and provided every evaluated state--action pair has positive representation in the logged data, the target policy can be reconstructed from reweighted behavior frequencies,
\begin{equation}
 \pi(a\mid s)=
 \frac{w_\pi(s,a)d_\beta(s,a)}{\sum_{a'}w_\pi(s,a')d_\beta(s,a')}.
 \label{eq:dicepolicy}
\end{equation}
Variants use this construction for off-policy evaluation, imitation, state-only or transition-only learning, mixed-quality demonstrations, constrained optimization, and cooperative multi-agent learning \citep{nachum2019dualdice,zhang2020gendice,kostrikov2020valuedice,lee2021optidice,lee2022coptidice,kim2022demodice,ma2022smodice,kim2022lobsdice,yu2023relaxdice,bui2025misodice}. They share density-ratio machinery but not one estimand.

For context $c$, demand $q_c$ converts episodic occupancy into expected link-turn flow, so DICE can evaluate a fixed policy from logs, imitate a demonstrator flow, or optimize a constrained policy where the logged data adequately represent the required states and actions. Activity choice has the same interpretation on a time--space--resource graph. The limits are strict: ratios are not taste coefficients, and data coverage must hold within the relevant destination, time, information, and traveler groups. Finite trips require episodic rather than stationary flow equations, while congestion makes single-agent ratios conditional on a fixed traffic regime \citep{nachum2019dualdice,lee2021optidice}.

\subsection{Learning from observations and incomplete trajectories}

Mobility records often omit actions, contain irregular samples, or leave path segments ambiguous. Rather than treating one map match as truth, Equation~\eqref{eq:fourlayer} can marginalize feasible latent paths or schedules. Recursive route likelihoods can integrate disconnected observations exactly, while state-transition occupancy methods learn when expert actions are absent \citep{torabi2018behavioral,sun2019observation,mai2023incomplete,pham2025iostom}. The same principle applies to coarse locations and missing activity episodes: uncertainty belongs in the observation model, not in a deterministic preprocessing decision.

\subsection{Learning from suboptimal, mixed-quality, and negative data}

Observed travel is not uniformly expert: mistakes, exploration, habits, and unsafe or inefficient behavior coexist. Quality-aware methods combine trusted, suboptimal, negative, and unlabeled demonstrations without assigning them equal authority \citep{hoang2024sprinql,hoang2025avoidance,bui2025misodice}. Transportation labels can reflect incidents, violations, detours, accessibility failures, or feedback, but must be tied to an explicit objective or constraint rather than a vague expert score.

\subsection{Offline reinforcement learning}

Offline RL optimizes a supplied return from fixed data and is therefore prescriptive, not a model of observed preference. Conservative methods limit exploitation of unsupported actions \citep{levine2020offline,kumar2020cql}. Transportation applications should keep traveler preference, the operator's safety/emissions/welfare objective, and recommendation acceptance separate; coverage, constraints, uncertainty, and human response determine whether prescription is credible \citep{lee2022coptidice}.

\subsection{Learning regime: offline, interactive, continual, and transfer}

``Online'' is used ambiguously in transportation: it can mean real-time prediction from a fixed model, sequential parameter updating as data arrive, or active interaction that changes which observations are collected. Only the last two are online learning. Most revealed mobility data support offline estimation because unrestricted exploration is often impractical or unethical in safety-, privacy-, or burden-sensitive applications. Table~\ref{tab:regimes} distinguishes offline, interactive, continual, and transfer regimes by their available information, transportation use, and principal validity risk.

\begin{longtable}{@{}P{0.18\textwidth}P{0.25\textwidth}P{0.26\textwidth}P{0.22\textwidth}@{}}
\caption{Learning regimes for sequential mobility choice.}\label{tab:regimes}\\
\toprule
\textbf{Regime} & \textbf{Information available} & \textbf{Mobility use} & \textbf{Principal validity risk} \\
\midrule
\endfirsthead
\toprule
\textbf{Regime} & \textbf{Information available} & \textbf{Mobility use} & \textbf{Principal validity risk} \\
\midrule
\endhead
\endfoot
\bottomrule
\endlastfoot
Offline estimation / IL \par\citep{ziebart2008maxent,ho2016gail} & \tabitem{Fixed trajectories, surveys, or transitions}\tabitem{No learner-induced data collection} & \tabitem{Describe observed behavior}\tabitem{Predict routes or schedules within the logged regime} & \tabitem{Coverage, self-selection, confounding, and unobserved availability}\tabitem{Do not interpret supported prediction as intervention evidence} \\
\mobilityrowrule
Offline RL / DICE \par\citep{levine2020offline,nachum2019dualdice,lee2021optidice} & \tabitem{Fixed logs plus a target policy, reward, or constraint}\tabitem{Known or estimated behavior support} & \tabitem{Off-policy evaluation and occupancy correction}\tabitem{Conservative or constrained recommendation} & \tabitem{Extrapolation when target actions lack support}\tabitem{Operator reward must remain separate from descriptive preference} \\
\mobilityrowrule
Interactive IL / active IRL \par\citep{ross2011dagger,biyik2018active,lindner2022active} & \tabitem{Expert corrections, trajectory comparisons, or controlled feedback}\tabitem{Queries selected by the learner} & \tabitem{Collect informative preference or constraint evidence}\tabitem{Correct errors in learner-visited states} & \tabitem{Query burden, safety, selection, and treatment effects}\tabitem{Report who was queried and how interaction changed behavior} \\
\mobilityrowrule
Continual learning \par\citep{parisi2019continual} & \tabitem{Time-ordered observations and repeated model updates}\tabitem{Versioned network and policy context} & \tabitem{Adapt to incidents, new services, and behavioral drift}\tabitem{Maintain real-time prediction performance} & \tabitem{Catastrophic forgetting and delayed labels}\tabitem{Separate temporary conditions from stable taste change} \\
\mobilityrowrule
Transfer / meta-learning \par\citep{finn2017maml,zhao2023deepirl,zhang2024metairl} & \tabitem{Related destinations, populations, periods, or networks}\tabitem{Small labeled target-domain sample} & \tabitem{Cold-start prediction and partial pooling}\tabitem{Estimate sparse contexts with shared structure} & \tabitem{Leakage through identities, geography, time, or repeated travelers}\tabitem{Declare which reward, representation, or observation component is invariant} \\
\end{longtable}

\subsection{What transportation applications currently establish}

Table~\ref{tab:evidence} summarizes transport-specific evidence rather than ranking methods. \rev{Among the studies reviewed, the clearest demonstrations concern network-scale recursive or inverse planning, context-dependent reward learning, inference from incomplete trajectories, and activity-schedule generation. Evidence is thinner for responses to actual interventions, preference--constraint separation, endogenous congestion or crowding feedback, and transfer across networks.} Unless explicitly marked ``emerging,'' each row reports a result evaluated in a transportation setting; \rev{the emerging tag identifies preprint, stylized, or narrowly evaluated evidence.}
\begin{longtable}{@{}P{0.18\textwidth}P{0.24\textwidth}P{0.27\textwidth}P{0.22\textwidth}@{}}
\caption{Representative transportation evidence for learned sequential choice.}\label{tab:evidence}\\
\toprule
\textbf{Study and domain} & \textbf{Data or setting} & \textbf{\rev{What was demonstrated or evaluated}} & \textbf{\rev{Remaining limitation or untested claim}} \\
\midrule
\endfirsthead
\toprule
\textbf{Study and domain} & \textbf{Data or setting} & \textbf{\rev{What was demonstrated or evaluated}} & \textbf{\rev{Remaining limitation or untested claim}} \\
\midrule
\endhead
\endfoot
\bottomrule
\endlastfoot
\citet{fosgerau2013link}, road routes & Network observations and recursive likelihood & \tabitem{Normalized probability over the full feasible path set}\tabitem{Likelihood evaluation without explicit path enumeration} & \tabitem{Utility remains parametric}\tabitem{Markov state, error distribution, and exogenous attributes are assumed} \\
\mobilityrowrule
\citet{zhao2023deepirl}, road routes & Shanghai taxi GPS trajectories & \tabitem{Nonlinear context-dependent reward representation}\tabitem{Prediction for destinations excluded from training} & \tabitem{Destination transfer does not prove reward invariance across cities or policies}\tabitem{No welfare-valid utility interpretation} \\
\mobilityrowrule
\textbf{Emerging:} \citet{barnes2023maps}, road routing & Industrial road graphs and large-scale inverse planning & \tabitem{Inverse planning on graphs with hundreds of millions of states}\tabitem{Graph compression and mixture-of-experts scalability} & \tabitem{Preprint evidence and proprietary evaluation}\tabitem{Observed routes are selected by a platform and need not reveal population preferences} \\
\mobilityrowrule
\citet{mai2023incomplete}, incomplete routes & Disconnected route observations & \tabitem{Exact marginalization over all feasible missing segments}\tabitem{Computational acceleration by decomposition--composition} & \tabitem{Depends on the specified route-choice and observation models}\tabitem{Does not address population selection or missing entire trips} \\
\mobilityrowrule
\citet{zimmermann2018activity}, activity schedules & Time--space activity network and survey observations & \tabitem{Joint recursive choice of activity, location, timing, and mode}\tabitem{Whole-schedule feasibility on an augmented network} & \tabitem{State and utility are analyst specified}\tabitem{Time/resource expansion creates computational growth} \\
\mobilityrowrule
\citet{drchal2019scheduler}, activity schedules & Regional synthetic population and travel evidence & \tabitem{Regional-scale daily schedule generation}\tabitem{Validation against activity-model population targets} & \tabitem{Supervised reproduction rather than reward recovery}\tabitem{Intervention and welfare validity were not established} \\
\mobilityrowrule
\citet{song2024stateirl}, activity travel & Cellular signaling trajectories & \tabitem{State-based IRL and reward-recovery experiments}\tabitem{Transfer tests across the evaluated settings} & \tabitem{Cellular states provide coarse activity semantics}\tabitem{Limited evidence under actual transport-policy interventions} \\
\mobilityrowrule
\citet{liu2025contextirl}, activity schedules & Large resident travel survey & \tabitem{Context-conditioned daily schedule generation}\tabitem{Generalization across held-out observed contexts} & \tabitem{One-day self-reported behavior limits longitudinal claims}\tabitem{MDP state, dynamics, and constraints remain assumed} \\
\mobilityrowrule
\citet{liang2026interpretable}, activity travel & Singapore travel-survey sequences & \tabitem{Policy distillation into readable decision rules}\tabitem{Short- and long-horizon reward diagnostics} & \tabitem{A faithful surrogate describes the learned policy, not necessarily true utility}\tabitem{Structural identification and intervention invariance remain open} \\
\mobilityrowrule
\citet{vo2026fusion}, multimodal demand & Survey and smart-card data in two cities & \tabitem{Cross-source recovery of time-dependent multimodal flows}\tabitem{Validation on data not used to fit the fusion model} & \tabitem{Aggregate flow agreement does not identify individual preferences}\tabitem{Person-level sequences and choice sets remain latent} \\
\mobilityrowrule
\citet{mo2026llmtravel}, prompted choice & Mode and trip-purpose prediction under label budgets & \tabitem{Competitive prediction when labeled examples are scarce}\tabitem{Comparison with baselines under matched label budgets} & \tabitem{Advantage narrows as supervised data increase}\tabitem{Prompted choices and rationales are not identified utilities} \\
\mobilityrowrule
\textbf{Emerging:} \citet{wang2025llmtraveler}, route adaptation & Stylized day-to-day congestion experiments & \tabitem{Memory-equipped agents reproduce selected route-switching patterns}\tabitem{Qualitative adaptation under repeated congestion feedback} & \tabitem{Stylized games and narrow behavioral targets}\tabitem{Model-version sensitivity and limited validation on observed field interventions} \\
\end{longtable}

The table deliberately does not count generic DICE variants, diffusion planners, scientific agents, or general multi-agent algorithms as established transportation evidence merely because their mathematical objects can be mapped to networks. Those methods enter the following section as transferable methods or emerging proposals until they are tested on transport-specific feasibility, population, transfer, or intervention outcomes.

\section{Where Machine Learning Adds Transportation Value}

ML is most useful when it resolves a transportation bottleneck in computation, behavioral representation, incomplete observation, or data integration. Generative, multi-agent, and large language model systems remain emerging unless evaluated for route or schedule feasibility, behavioral interpretation, network response, and transfer across transportation settings. The discussion is organized around four transportation needs, each paired with its behavioral requirement.

\subsection{Network-scale computation and behavioral representation}

Sparse linear algebra, graph compression, warm starts, batching, and learned value approximations reused across planning queries can reduce computation over large graphs and activity states. Automatic differentiation enables gradient-based estimation through compatible solvers \citep{baydin2018autodiff,paszke2019pytorch,barnes2023maps}. Approximation can alter likelihoods and policies, so studies should report runtime, memory, planner calls, precision, solver tolerance, and Bellman or flow residuals. Sample efficiency must be measured in scarce independent units--travelers, days, destinations, cities, interventions, labels, or simulator calls--with leakage-resistant splits and equal target-domain information \citep{finn2017maml,biyik2018active,zhang2024metairl}. Standard ML controls such as fixed tuning budgets, ablations, calibration, uncertainty, and repeated seeds complement, rather than replace, behavioral tests \citep{guo2017calibration,lakshminarayanan2017deepensembles,henderson2018matters}.

Linear-in-parameters utility is interpretable but can miss nonlinear interactions among time, road environment, weather, trip purpose, schedule pressure, and personal constraints. A useful hybrid is
\begin{equation}
 r(s,a,c)=\boldsymbol{\beta}^{\top}\mathbf{x}(s,a,c)
 +g_\phi(\mathbf{z}(s,a,c)),
 \label{eq:hybridreward}
\end{equation}
where $\mathbf{x}$ contains policy-relevant attributes with sign, scale, or monotonicity restrictions, and $g_\phi$ is a regularized learned residual. Under the required scale, exogeneity, and choice-set assumptions, the interpretable term can support values of time and elasticities; the residual captures additional interactions and perceptual context. Orthogonality penalties, residual centering, or staged estimation can prevent the flexible residual from reproducing and obscuring the interpretable component \citep{zhao2023deepirl,liang2026interpretable}.

Heterogeneity should also be structured. Hierarchical or latent-class rewards can distinguish stable taste variation; embeddings can capture remaining high-dimensional variation; and random scale should not be confused with random preferences. Sharing information across travelers while retaining individual differences is preferable to fitting one unconstrained model per traveler, especially when personal histories are short \citep{train2009discrete}.

Graph networks can encode topology, land use, turns, transfers, and neighborhood context as features, rewards, or value corrections inside a solver that explicitly enforces the declared action set \citep{battaglia2018relational}. Unseen-region and unseen-network tests, identifier checks, and topology changes are needed to distinguish transfer from memorization.

Transformers can represent activity histories, household context, and non-Markovian route dependence \citep{chen2021decision}. Feasible-action masks, time/resource states, graph positional encodings, and terminal-consistency losses must preserve connectivity and schedule logic. Sequence prediction or return conditioning does not by itself identify utility.

\subsection{Feasible generation, constraints, risk, and uncertainty}

Diffusion and flow-matching models can represent multimodal route or schedule distributions and assist map matching, trajectory completion, and simulation \citep{janner2022diffuser,chi2023diffusion,lipman2023flow}. They should decode over feasible paths or generate latent preference representations for a feasibility-enforcing solver; otherwise generated outputs may violate connectivity, transfers, or time budgets. Calibration, rare-event coverage, controllability, and intervention response remain open tests.

Travel behavior contains restrictions that should not all be represented as large negative utilities. A context-specific constrained MDP (CMDP) augments the reward with constraint-cost functions $\kappa_j(s,a,c)$ and budgets $b_j(c)$. For an episodic or discounted mobility process, it solves
\begin{equation}
 \begin{aligned}
 \max_{\pi}\quad
 J_r(\pi\mid c)
 &=\mathbb{E}_{\pi,P}\!\left[\sum_{t=0}^{T-1}\gamma^t r(s_t,a_t,c)\right],\\
 \text{s.t.}\quad
 J_{\kappa_j}(\pi\mid c)
 &=\mathbb{E}_{\pi,P}\!\left[\sum_{t=0}^{T-1}\gamma^t
 \kappa_j(s_t,a_t,c)\right]\le b_j(c),
 \quad j=1,\ldots,m .
 \end{aligned}
 \label{eq:cmdp}
\end{equation}
CMDPs formalize cumulative limits \citep{altman1999cmdp}; safe RL adds safety filters that can override unsafe actions, robust backups, and constraint-optimization methods \citep{garcia2015safe}. Costs can represent expenditure, risk, emissions, missed commitments, or battery depletion. Known physical and legal rules instead belong in $\mathcal{A}(s,c)$ or the transition kernel: a finite penalty wrongly leaves an impossible action available. Thus availability determines what can be chosen, utility ranks feasible choices, and a CMDP couples choices through a trajectory-level budget.

\emph{Constrained IRL} learns reward while enforcing known constraints \citep{ding2022xmen}; \emph{inverse constraint learning} infers constraint costs given a known or separately specified reward \citep{malik2021icrl,quan2024icsdice}. Jointly learning both is weakly identified because an unchosen link or activity may reflect low utility, nonavailability, a personal constraint, missing information, or absent data support.

Regulations, schedules, accessibility audits, vehicle characteristics, commitments, stated availability, multiple tasks, and negative examples can anchor this distinction \citep{kim2023sharedconstraints,hoang2024goodbad}. Studies should state which constraints are known, inferred, or assumed; their units and heterogeneity; and the sensitivity of reward estimates to alternative specifications.

\begin{longtable}{@{}P{0.17\textwidth}P{0.22\textwidth}P{0.25\textwidth}P{0.25\textwidth}@{}}
\caption{Constraint roles in sequential mobility choice.}\label{tab:constraints}\\
\toprule
\textbf{Constraint role} & \textbf{Choice-model meaning} & \textbf{Mobility examples} & \textbf{Recommended treatment} \\
\midrule
\endfirsthead
\toprule
\textbf{Constraint role} & \textbf{Choice-model meaning} & \textbf{Mobility examples} & \textbf{Recommended treatment} \\
\midrule
\endhead
\endfoot
\bottomrule
\endlastfoot
Known physical or legal & \tabitem{Determines whether an alternative exists}\tabitem{Rules out impossible state transitions} & \tabitem{Road connectivity, closures, one-way rules}\tabitem{Service hours, mode eligibility, and inaccessible facilities} & \tabitem{Encode exactly in the action mask, transition kernel, or schedule network}\tabitem{Do not approximate a known prohibition with a finite reward penalty} \\
\mobilityrowrule
Known cumulative & \tabitem{Couples current actions to a finite trajectory-level resource}\tabitem{May bind only after several decisions} & \tabitem{Daily time or fare budget}\tabitem{Battery range, duty hours, transfer limits, and emissions cap} & \tabitem{Track the resource in the state or impose a CMDP expectation/chance constraint}\tabitem{Report units, budget, and whether violation is hard or probabilistic} \\
\mobilityrowrule
Latent traveler or household & \tabitem{Unobserved capability, availability, or commitment}\tabitem{Varies across people or episodes} & \tabitem{Mobility limitation, childcare, or vehicle access}\tabitem{Personal safety threshold and mandatory activity} & \tabitem{Use hierarchical latent constraints or inverse constraint learning}\tabitem{Anchor with surveys, repeated tasks, regulations, or negative examples and test reward sensitivity} \\
\mobilityrowrule
Operator or social & \tabitem{Limits a recommended system policy}\tabitem{Represents the planner's objective, not revealed traveler preference} & \tabitem{Safety exposure, fleet emissions, and service equity}\tabitem{Capacity, subsidy, or operating budget} & \tabitem{Use constrained offline RL or optimization}\tabitem{Add traveler acceptance and equilibrium response before system evaluation} \\
\end{longtable}

Operator constraints are prescriptive and should remain separate from traveler utility, acceptance, and inferred personal limitations. Constraint satisfaction alone does not establish behavioral validity.

Risk sensitivity and robustness address different sources of uncertainty. Conditional value at risk (CVaR) or distributional returns may represent a traveler's or operator's criterion over uncertain outcomes, whereas robust MDPs commonly guard against ambiguity in transition models \citep{iyengar2005robust,nilim2005robust,chow2015risk}. Studies should therefore state whose risk criterion is modeled and distinguish outcome risk from model uncertainty; otherwise cautious predictions may be misinterpreted as traveler risk aversion.

\subsection{Heterogeneous mobility data and privacy}

Surveys provide semantics and constraints but few people or days; passive traces provide scale but suffer selection, missing purpose, variable resolution, and incomplete modes; counts and census data add aggregate states or margins rather than individual choices. Fusion must preserve these different units, supports, and errors \citep{huang2018fusion,graellsgarrido2023fusion,vo2026fusion}.

The four-layer model supplies a principled formulation. Let $z^{(m)}$ indicate inclusion in source $m$, $y^{(m)}$ denote its observed record, $\mathcal{H}_m$ its spatial, temporal, semantic, and aggregation operator, and $q_m$ source-quality variables such as sampling interval, positional accuracy, penetration rate, or survey weight. Conditional on a latent trajectory $\tau$,
\begin{equation}
 \begin{aligned}
 p(\{z^{(m)},y^{(m)}\}_{m=1}^{M}\mid c)
 &=\int p_\theta(\tau\mid c)
 \prod_{m=1}^{M}p_{\eta_m}\!\left(z^{(m)}\mid\tau,c,q_m\right)\\[-0.1em]
 &\quad\times\left[p_{\psi_m}\!\left(y^{(m)}\mid
 \mathcal{H}_m(\tau,c),q_m,z^{(m)}=1\right)\right]^{z^{(m)}}d\tau .
 \end{aligned}
 \label{eq:datafusion}
\end{equation}
The operator $\mathcal{H}_m$ maps a latent route or schedule to each source's spatial, temporal, semantic, and aggregation scale; $p_{\eta_m}$ separates measurement error from inclusion driven by ownership, use, response, provider coverage, or device settings. Residual dependence across sources requires a shared selection variable or joint source model. This formulation lets sparse surveys identify cross-modal relationships while dense passive data anchor time-dependent flows \citep{huang2018fusion,graellsgarrido2023fusion,vo2026fusion}.

\begin{longtable}{@{}P{0.19\textwidth}P{0.33\textwidth}P{0.36\textwidth}@{}}
\caption{Complementary roles and failure modes in behavioral data fusion.}\label{tab:datafusion}\\
\toprule
\textbf{Source} & \textbf{Information contributed} & \textbf{Source-specific limitation to model} \\
\midrule
\endfirsthead
\toprule
\textbf{Source} & \textbf{Information contributed} & \textbf{Source-specific limitation to model} \\
\midrule
\endhead
\endfoot
\bottomrule
\endlastfoot
Survey / activity diary & \tabitem{Trip purpose, attitudes, household roles, and reported constraints}\tabitem{Socioeconomics, available modes, and stated/revealed alternatives} & \tabitem{Few observation days and respondents}\tabitem{Recall error, nonresponse, survey design effects, and uncertain expansion weights} \\
\mobilityrowrule
GPS / smartphone panel & \tabitem{Fine route geometry and departure/arrival timing}\tabitem{Repeated within-person routes and day-to-day adaptation} & \tabitem{Selective participation and device-dependent sampling}\tabitem{Battery gaps, map-match ambiguity, missing modes, and weak activity semantics} \\
\mobilityrowrule
Smart card / ticketing & \tabitem{Dense longitudinal boarding and transfer transactions}\tabitem{Service-specific timing and aggregate transit demand} & \tabitem{Transit-user population only}\tabitem{Missing alighting/access legs, card sharing, fare-policy censoring, and uncertain trip purpose} \\
\mobilityrowrule
Cellular / location service & \tabitem{Broad spatial coverage and long observation windows}\tabitem{Population-scale origin--destination and activity-location signals} & \tabitem{Coarse, irregular, and device-dependent resolution}\tabitem{Provider selection, uncertain stops, weak semantics, and changing tower/app coverage} \\
\mobilityrowrule
Traffic counts / probes & \tabitem{Link flow, speed, congestion, reliability, and incident response}\tabitem{External aggregate constraints for network loading} & \tabitem{No traveler identity, purpose, or perceived choice set}\tabitem{Incomplete trajectories and possible probe-fleet/platform selection} \\
\mobilityrowrule
Census / time use / land use & \tabitem{Population and household margins}\tabitem{Activity-duration norms, opportunities, accessibility, and land-use context} & \tabitem{Aggregated and infrequently updated}\tabitem{Not directly linked to individual trajectories; spatial/temporal scales may mismatch decisions} \\
\end{longtable}

Record-level fusion propagates linkage uncertainty; representation-level fusion combines source-specific encoders; population-level fusion aligns unlinked samples through shared parameters or margins; and prediction-level fusion ensembles posteriors when linkage is unavailable \citep{graellsgarrido2023fusion,bai2024multiresolution}. Likelihood scaling, survey weights, noise models, or partial pooling should prevent a massive low-detail source from overwhelming a small semantically rich one. Credibility requires leave-one-source-out tests, independent-survey validation, calibration across demographic and resolution strata, linkage sensitivity, and individual- and aggregate-level posterior checks.

Because linked mobility traces are highly identifying, studies should minimize person-level linkage, use secure or privacy-preserving computation where needed, and report authority, privacy parameters, utility loss, and subgroup effects \citep{dwork2014privacy}.

\subsection{System interaction and semantic tools}

Strategic models are needed when choices alter others' actions or payoffs through congestion, crowding, coordination, platforms, or automated agents; heterogeneity alone can remain hierarchical. Named-agent games suit households and fleets, while mean fields or assignment better represent anonymous congestion \citep{aguirregabiria2007games,yang2018meanfield,ameli2022departure,shou2022markovrouting,yu2019maairl}. Cooperative occupancy methods do not resolve competitive rewards or equilibrium selection \citep{bui2025misodice}.

Estimation must separate preferences, information, interaction, and solution concept. Policy counterfactuals should recompute all affected policies and network states and report equilibrium residuals, congestion, welfare, population sensitivity, and distributional effects. Fixed-regime single-agent predictions cannot support such claims; mixed human--automated simulations remain emerging evidence \citep{akman2025routerl}.

LLMs are emerging semantic tools, not established behavioral estimators. They can extract purposes or constraints from text and propose high-level plans, but outputs should use structured fields with predefined names and permitted values, retain source links, be audited, and be passed to calibrated probabilistic choice models or explicit solvers. Low-data prediction and stylized agent simulations are promising, yet do not establish preferences, elasticities, population shares, or intervention validity \citep{mo2026llmtravel,wang2025llmtraveler,liu2026gatsim,santos2026citybehavex}.

Studies should minimize identifying text; archive prompts and model versions; report variation, cost, and schema failures; compare equal-information non-LLM baselines; and validate synthetic schedules against population margins and held-out behavior. Generated rationales and research-agent proposals require executable verification \citep{trinh2024alphageometry,romeraparedes2024funsearch}.

Approximate solvers, minibatching, mixed precision, pretraining, fusion, and synthetic generation can introduce numerical bias, selection, leakage, privacy loss, and reproducibility or energy costs. Direct diagnostics include solver and flow residuals, propagation of approximation error, tests separating taste from scale and availability, negative-transfer reporting, and learning curves based on independent travelers or interventions. Predictive benchmarks should be paired with behavioral validity tests and full reporting of sampling corrections, tuning, hardware, precision, energy, and solver tolerances.

\section{A Behaviorally Disciplined Modeling Workflow}

The modeling prescription follows directly: retain an explicit state, feasible actions, resource constraints, reward normalization, observation and selection processes, and equilibrium feedback. Attach learned components only to declared representation, data, or computation bottlenecks \citep{mai2020relation,train2009discrete,sifringer2020representation,han2022tastenet}. Every component should state its estimand, inputs, constraints, approximation, and defensible claim. Calibrated prediction cannot repair a misspecified observation model, and identified utility cannot support a congestion counterfactual without system feedback.
The following nine-step workflow turns that principle into an implementable model design. The steps are ordered from the behavioral data-generating process to system-level validation; each step should produce an explicit modeling object and a diagnostic that can fail.
\begin{enumerate}
\item \textbf{Construct a decision state.} Define the information available when the traveler acts: location, destination or activity purpose, clock time, network conditions, remaining resources, commitments, and any necessary memory. \rev{Compare nested state definitions and test whether omitted history systematically predicts the next action; persistent predictive value is evidence against, rather than a conclusive test of, Markov sufficiency.}

\item \textbf{Separate feasibility, constraints, and preference.} Put physically or institutionally impossible actions in the feasible set $\mathcal{A}(s,c)$ and transition model $P$, and represent cumulative limits such as time, money, range, or mandatory activities through resource states or a constrained MDP. Infer a latent constraint only when repeated variation can distinguish an unavailable action from an available but unattractive one.

\item \textbf{\rev{Specify an interpretable reward backbone.}} Include observed, policy-relevant attributes such as time, monetary cost, reliability, schedule delay, and accessibility before adding flexible residual terms. \rev{Fix the utility location and scale normalizations, impose defensible signs or monotonicity where appropriate, and state which coefficients are intended to support elasticities, value-of-time, or welfare calculations.}

\item \textbf{\rev{Add a scope-restricted learned component.}} Use graph, sequence, or multimodal encoders only for a declared bottleneck, such as spatial context, long history, unobserved heterogeneity, or costly value-function approximation. Restrict the component through architecture, regularization, or residualization, and report whether it changes interpretable coefficients, calibration, transfer, and computation relative to the backbone.

\item \textbf{Use LLMs at semantic boundaries.} \rev{Let an LLM extract purposes, constraints, or high-level plans from text only through structured fields with predefined names and values, source provenance, versioned prompts, and an explicit treatment of uncertain or invalid outputs.} Pass those outputs to a calibrated choice model or explicit planner, and measure formatting failures and downstream sensitivity rather than treating fluent text as a behavioral decision rule.

\item \textbf{Match the learning objective to the use case.} Use recursive choice when normalized probabilities and interpretable trade-offs are central, IRL when a transferable reward is the estimand, occupancy methods for distributional imitation or simulation, behavior cloning for supported prediction, and offline RL for a supplied prescriptive objective. State explicitly whether the output is utility, reward, policy, occupancy, constraint, or trajectory distribution.

\item \textbf{Model observation and selection.} Connect the latent trajectory to GPS, diary, smart-card, or cellular records through source-specific error and missingness models. Marginalize uncertain paths or episodes, model who and what enters each source, and retain multiple latent explanations whenever deterministic preprocessing would understate uncertainty.

\item \textbf{Resolve system feedback.} Load predicted choices onto the network and iterate with congestion, crowding, service capacity, or accessibility until the stated assignment or equilibrium condition is satisfied. Keep individual utility separate from the operator's objective, and report convergence residuals and distributional system outcomes.

\item \textbf{\rev{Evaluate evidence at the level of the claim.}} Apply the claim-specific tests in Section~\ref{sec:evaluation}, from calibration and complete-sequence feasibility to behavioral validity, transfer, and intervention response. Prediction alone does not establish preference recovery, transfer, or policy validity.
\end{enumerate}

\begin{longtable}{@{}P{0.16\textwidth}P{0.24\textwidth}P{0.28\textwidth}P{0.22\textwidth}@{}}
\caption{Recommended model choice by scientific objective.}\label{tab:decision}\\
\toprule
\textbf{Objective} & \textbf{\rev{Preferred modeling core}} & \textbf{Useful ML augmentation} & \textbf{Minimum evidence before use} \\
\midrule
\endfirsthead
\toprule
\textbf{Objective} & \textbf{\rev{Preferred modeling core}} & \textbf{Useful ML augmentation} & \textbf{Minimum evidence before use} \\
\midrule
\endhead
\endfoot
\bottomrule
\endlastfoot
Behavioral explanation & \tabitem{Recursive or dynamic discrete choice}\tabitem{Explicit choice set, normalization, and observation model} & \tabitem{Structured nonlinear residual}\tabitem{Hierarchical heterogeneity or learned context representation} & \tabitem{Signs, scale, elasticities, and substitution tests}\tabitem{Sensitivity to availability, endogeneity, and observation error} \\
\mobilityrowrule
\rev{New-destination generalization} & \tabitem{IRL with a destination-conditioned reward}\tabitem{\rev{Explicit network transitions and replanning}} & \tabitem{Graph reward encoder}\tabitem{Amortized value function or meta-learning initialization} & \tabitem{Destinations and nearby links excluded without leakage}\tabitem{Reward contrasts and performance stable across destination types} \\
\mobilityrowrule
Population simulation & \tabitem{Calibrated probabilistic choice or occupancy matching}\tabitem{Population weights and network loading} & \tabitem{Graph/sequence policy with action masks}\tabitem{Generative heterogeneity and fast rollout} & \tabitem{Joint route/schedule distributions and population moments}\tabitem{Complete-sequence feasibility, calibration, and congestion consistency} \\
\mobilityrowrule
Missing-trajectory recovery & \tabitem{Latent path or schedule likelihood}\tabitem{Source-specific sensor and missingness model} & \tabitem{Generative proposal or observation encoder}\tabitem{Probabilistic map matching and amortized inference} & \tabitem{Synthetic masking plus real disconnected records}\tabitem{Calibration of path uncertainty and checks against exact marginalization} \\
\mobilityrowrule
Recommen\-dation & \tabitem{Separate preference and acceptance models}\tabitem{Constrained offline RL for the operator objective} & \tabitem{Conservative value learning}\tabitem{Uncertainty, safety shield, and personalized acceptance} & \tabitem{Action-level coverage and off-policy validation}\tabitem{Safety/resource compliance and human response in staged deployment} \\
\mobilityrowrule
Policy counterfactual & \tabitem{\rev{Interpretable reward with explicit constraints and observation process}}\tabitem{Assignment, mean-field, or game equilibrium} & \tabitem{Differentiable solver and domain adaptation}\tabitem{Sensitivity or partial-identification bounds} & \tabitem{Pre/post intervention prediction without refitting}\tabitem{Equilibrium residuals, welfare stability, and distributional impacts} \\
\end{longtable}

\rev{This architecture clarifies when simpler methods may be preferable. If the network is modest, a structured specification passes fit, calibration, and substitution tests, and policy interpretation is central, recursive logit may be preferable to a deep model.} If only next-link prediction is required in a stable environment, masked behavior cloning may be sufficient. Complexity is justified by a demonstrated gain in scale, transfer, data integration, or specification robustness--not by model class alone.

\section{Evaluation and Benchmarking}\label{sec:evaluation}

\rev{Evaluation must match the intended claim, because methods that estimate different objects cannot be ranked responsibly by one accuracy measure. Evidence progresses from held-out probability and trajectory quality, through complete-sequence feasibility and behavioral validity, to recovery, transfer, intervention response, and system outcomes. Recovery tests use data with known generating parameters; generalization tests hold out people, destinations, networks, or periods; transfer tests introduce a substantively different domain; and intervention tests change the policy environment. In the running commute, next-link accuracy, recovery of time and cost trade-offs, complete-day feasibility, and post-toll flow response are therefore distinct tests. Each study should report every layer required by its strongest claim.}

\rev{Random trip-level splits are often inadequate because the same traveler, destination, and nearby links can appear in training and test data. We therefore propose four complementary benchmark levels. The first two use (i) synthetic networks with known rewards, constraints, and observation noise and (ii) a public road or transit network with map-matched trajectories. The next two use (iii) an activity-diary or mobility dataset with time, purpose, and personal context and (iv) an intervention or domain shift involving closures, tolls, incidents, service changes, or a new city} \citep{arora2021survey,levine2020offline,zhao2023deepirl}. \rev{Individual studies need only use the levels required by their claims, while a community benchmark would ideally cover all four.}

The core reporting suite should cover probabilistic quality, trajectory fidelity, behavioral validity, robustness, computation, and reproducibility. Recovery, transfer, counterfactual, online, data-fusion, safety, system, privacy, and LLM tests are claim-dependent and should be added whenever the corresponding capability is asserted.

\begin{longtable}{@{}P{0.18\textwidth}P{0.34\textwidth}P{0.34\textwidth}@{}}
\caption{Recommended evaluation suite for learned travel behavior.}\label{tab:benchmark}\\
\toprule
\textbf{Dimension} & \textbf{Concrete measures and test design} & \textbf{\rev{Why the measure matters / interpretation of failure}} \\
\midrule
\endfirsthead
\toprule
\textbf{Dimension} & \textbf{Concrete measures and test design} & \textbf{\rev{Why the measure matters / interpretation of failure}} \\
\midrule
\endhead
\endfoot
\bottomrule
\endlastfoot
\rev{Probabilistic predictive quality} & \tabitem{Held-out negative log likelihood or cross-entropy}\tabitem{Reliability diagrams, expected calibration error, and top-$k$ coverage} & \tabitem{A plausible modal prediction may hide a badly estimated probability distribution}\tabitem{Downstream sampling and uncertainty estimates are unreliable} \\
\mobilityrowrule
Trajectory fidelity & \tabitem{Edge/episode overlap, edit or dynamic-time-warping distance}\tabitem{Destination completion and joint timing, duration, and chaining statistics} & \tabitem{High next-step accuracy may still produce disconnected paths or incoherent days}\tabitem{Sequence dependence is not represented adequately} \\
\mobilityrowrule
Behavioral validity & \tabitem{Coefficient signs, value of time, direct/cross elasticities, and monotonicity}\tabitem{Link-duplication, overlap, and alternative-availability substitution tests} & \tabitem{The model uses a predictive shortcut or implausible trade-off}\tabitem{\rev{Good fit alone does not support welfare or policy interpretation}} \\
\mobilityrowrule
Recovery & \tabitem{Known-truth synthetic tests for utility, reward ranking, and constraints}\tabitem{Shaping-invariant induced-policy and occupancy comparisons} & \tabitem{Different latent explanations produce the same behavior}\tabitem{Reported reward or constraint is not identified by the design} \\
\mobilityrowrule
Transfer & \tabitem{Hold out destinations, travelers, periods, regions, and complete networks}\tabitem{Report both source and target performance plus negative-transfer cases} & \tabitem{The representation memorizes identifiers or local opportunity structure}\tabitem{Claimed reward/policy invariance does not survive domain shift} \\
\mobilityrowrule
Sample efficiency & \tabitem{Learning curves over independent travelers, days, labels, queries, and planner calls}\tabitem{Equal target-domain information, tuning budget, and pretraining disclosure} & \tabitem{Repeated correlated transitions exaggerate effective sample size}\tabitem{Improvement may result from leakage or unequal information budgets} \\
\mobilityrowrule
Counter\-factual validity & \tabitem{Fit pre-intervention; predict post-toll, closure, service, or accessibility outcomes without refitting}\tabitem{Compare likelihood, flow change, elasticity, welfare, and equilibrium residuals} & \tabitem{\rev{Status-quo associations may not generalize to an intervention; predictive success does not by itself identify causal welfare effects}}\tabitem{Fixed-regime predictions omit behavioral or network adaptation} \\
\mobilityrowrule
Online / continual validity & \tabitem{One-step-ahead loss as observations arrive, adaptation delay, forgetting, and parameter drift}\tabitem{Query burden, safety events, and intervention exposure} & \tabitem{Updating is unstable, unsafe, or too costly}\tabitem{Temporary network change is being mistaken for permanent preference drift} \\
\mobilityrowrule
Robustness & \tabitem{Vary GPS noise, sampling interval, missing segments, and map-match ambiguity}\tabitem{Vary expert quality, label errors, transition uncertainty, and state specification} & \tabitem{Performance depends on one preprocessing path or optimal-expert assumption}\tabitem{Uncertainty has been hidden rather than propagated} \\
\mobilityrowrule
Fusion validity & \tabitem{Leave-one-source-out tests and independent-source validation}\tabitem{Sensitivity to source weights, record linkage, resolution, and selection models} & \tabitem{A large weak source overwhelms semantically rich data}\tabitem{Sampling frames or observation units are incompatible} \\
\mobilityrowrule
Feasibility / safety & \tabitem{Invalid transitions, missed obligations, and time/resource violations}\tabitem{Safety-event rate, CVaR, worst-case outcome, and safety-filter overrides} & \tabitem{Generated behavior is operationally unusable even if statistically plausible}\tabitem{Soft penalties are not enforcing hard rules} \\
\mobilityrowrule
System impact & \tabitem{Congestion, crowding, reliability, emissions, welfare, and accessibility}\tabitem{Distribution by income, geography, mode access, and traveler type} & \tabitem{Individually accurate choices can aggregate to harmful or unstable outcomes}\tabitem{Benefits and burdens may be distributed inequitably} \\
\mobilityrowrule
Computation & \tabitem{Wall-clock time, peak memory, planner calls, precomputation, latency, and energy}\tabitem{Hardware, batch size, numerical precision, tolerance, and residuals} & \tabitem{A scalability claim cannot be reproduced or compared}\tabitem{Approximation error may alter likelihoods, policies, or equilibria} \\
\mobilityrowrule
Reproduc\-ibility & \tabitem{Repeated seeds, uncertainty intervals, and clustered statistical errors}\tabitem{Code/data versions, tuning budget, software, hardware, and precision} & \tabitem{Reported gain may be a favorable random seed or unequal tuning effort}\tabitem{Result cannot be independently reconstructed} \\
\mobilityrowrule
Privacy & \tabitem{Linkage authority, privacy budget, attack or re-identification tests}\tabitem{Predictive and behavioral utility loss overall and by subgroup} & \tabitem{Fusion may expose sensitive routines or identities}\tabitem{Privacy protection may degrade rare-route or minority-group performance disproportionately} \\
\mobilityrowrule
LLM reliability & \tabitem{Prompt, model and corpus versions; repeated-run variance; schema validity}\tabitem{Rationale faithfulness, equal-information baselines, token/energy cost, and solver verification} & \tabitem{Semantic variables or agent choices are version-sensitive and non-reproducible}\tabitem{Fluent rationales may not explain the generated decision} \\
\end{longtable}

Comparisons should control the information set, feasible action space, target-domain sample, and tuning budget. A neural model given destination, map imagery, and personal history should not be compared to a linear choice model lacking those inputs and described as intrinsically superior. \rev{Candidate baselines, selected according to the application and claim, include shortest path, path-based logit with sampling correction, recursive logit, a flexible supervised policy, and the proposed behaviorally disciplined hybrid with ablations.} Learning curves should use independent travelers or days rather than treating correlated link transitions as new samples. Stochastic pipelines should report repeated seeds and uncertainty intervals, while statistical uncertainty should include traveler or origin--destination clustering where appropriate \citep{henderson2018matters}.

\rev{A model claimed to support intervention prediction should be fitted before an intervention and evaluated on subsequent route, mode, timing, or activity changes without refitting.} When no natural intervention is available, synthetic and semi-synthetic tests can vary network attributes while keeping known preferences fixed. \rev{Such tests assess intervention generalization; additional causal and welfare assumptions are needed for causal policy interpretation.} Sensitivity analyses should show how results change with reward normalization, transition uncertainty, missing-data assumptions, and equilibrium specification.

\section{\rev{Implications for Transportation-System Research}}

\rev{For travel-demand modeling, the principal implication is that route and activity trajectories should not be treated as interchangeable training sequences. Route models require network connectivity, destination completion, and substitution tests; activity models additionally require timing, duration, commitments, and resource accounting. The learned output must also match the planning use. A next-action policy may support short-horizon prediction, whereas value-of-time, welfare, and intervention claims require identified trade-offs, an observation model, and the relevant equilibrium response.}

\rev{For emerging mobility data and artificial intelligence (AI), the transportation contribution lies in the system insight enabled by the technology rather than in the technology alone. Graph and sequence representations can expand context and scale; passive traces can improve temporal and spatial coverage; and generative or language models can assist completion and semantic extraction. Their value should therefore be demonstrated through transportation outcomes such as complete-path probability, schedule feasibility, calibrated demand, congestion response, accessibility, reliability, or prediction after an intervention without refitting. Source selection, privacy, and observation error remain part of the behavioral model rather than preprocessing details.}

\rev{For network management and policy analysis, individual predictions must be loaded back into the transportation system. Tolls, closures, information, new services, and automated agents can change both traveler incentives and the congestion or crowding experienced by others. Consequently, policy-facing studies should recompute the relevant assignment, mean-field, or game equilibrium and report system performance and distributional outcomes. Shared datasets, leakage-resistant splits, solver residuals, and intervention tests would make these results more comparable across the transportation research community.}

\section{Research Agenda}
\rev{The preceding synthesis identifies a gap between what current models predict and what transportation policy applications require. Five priorities address whether a learned object remains interpretable, feasible, and reliable when the population, network, data source, or policy changes.}

\rev{First, equivalence results should state the conditions under which recursive logit, maximum causal entropy policies, dynamic logit, global path distributions, and regularized flows coincide, including the treatment of cycles, discounting, heterogeneous scale, stochastic transitions, and partial observation.} Companion identification analyses should separate what is recoverable under reward shaping, omitted constraints, learned states, and endogenous network conditions \citep{mai2020relation,ng1999shaping}.

Second, activity choice should become a first-class sequential-learning benchmark. An open time--space environment should include duration, household coordination, accessibility, resource constraints, and distinct survey and passive-data observation models. Success should require complete-day feasibility, population moments, and response to schedule or accessibility interventions, not only next-activity accuracy \citep{song2024stateirl,liu2025contextirl}.

\rev{Third, generalization and transfer should be tested layer by layer. Destinations and time periods can assess within-environment generalization, while changes in networks, populations, observation systems, or policies can assess transfer. These dimensions should be varied separately, with prespecified target-domain performance and parameter-stability criteria for rewards, policies, representations, observation models, or equilibrium parameters} \citep{zhao2023deepirl,zhang2024metairl}. Solver-aware representations should accompany these tests with declared approximation error, memory, planner calls, hardware, and energy \citep{barnes2023maps,mai2023incomplete}.

Fourth, preference, constraint, belief, observation, and equilibrium effects should be separated using repeated tasks, regulations, stated choices, incident information, and sensor-quality measures. Partial-identification or sensitivity bounds may be more credible than one recovered reward--constraint surface \citep{malik2021icrl,quan2024icsdice}. Policy experiments must recompute congestion, crowding, and acceptance rather than replaying a policy fitted under the old regime \citep{ameli2022departure,shou2022markovrouting}.

\rev{Fifth, the field needs a shared benchmarking protocol with common networks and intervention splits, source-aware survey and passive-data benchmarks, and interoperable likelihood and planning interfaces. Studies should report normalizations, solver tolerances, tuning budgets, uncertainty, and failed behavioral-validity tests} \citep{henderson2018matters,graellsgarrido2023fusion,vo2026fusion}. LLM-assisted variables or plans should enter only through versioned, audited, and solver-verified workflows. \rev{Near-term work should establish these controls before city-scale multi-agent or generative systems are used for behavioral policy analysis.}

\section{Limitations of the Review}

This review is purposive and integrative. It does not provide exhaustive retrieval, a registered protocol, duplicate independent screening, or PRISMA-style counts, and relevant studies may therefore have been missed. Its evidence labels are qualitative judgments about the strongest supported claim rather than meta-analytic effect estimates. The underlying studies also differ substantially in networks, data sources, baselines, and evaluation units, precluding quantitative aggregation.

The literature is moving quickly, particularly for LLM agents, generative trajectories, offline learning, and 2025--2026 transportation applications. Preprints and early studies are retained only to describe scale or frontier directions and are marked as emerging, but their conclusions may change. The framework also privileges sequential state--action representations; qualitative accounts, psychological process models, habitual behavior, and institutional explanations receive less attention unless they can be connected to a sequential choice state, belief, or constraint.

Finally, the authors' research overlaps with several reviewed areas, including recursive route choice, incomplete observations, and learning from heterogeneous demonstrations. To reduce the risk of overinterpretation, the synthesis distinguishes transport evidence from transferable methods, states the boundary of each cited result, and avoids numerical prevalence claims. A future systematic review with an independently coded public corpus would be a useful complement to the conceptual synthesis offered here.

\section{Conclusion}
\rev{Route and activity choice can share a sequential mobility framework without becoming the same empirical problem. Under stated assumptions, soft-Bellman and occupancy-flow connections provide a common mathematical language for recursive choice, IRL, IL, and network flow. Their learned objects nevertheless retain different meanings: a policy predicts behavior, a reward ranks trajectories under a specified model, utility coefficients may support behavioral trade-offs under identification assumptions, and system counterfactuals additionally require congestion, crowding, or equilibrium response. The practical conclusion for transportation research is therefore to match each learned object to its planning use and evaluate every layer required by that use. Progress depends less on another predictive architecture than on a stronger evidentiary chain: feasible complete sequences, calibrated uncertainty, interpretable substitution, transparent observation models, transfer across genuinely different networks, and credible evaluation after transportation interventions.}

\section*{CRediT Authorship Contribution Statement}
Hung Tran: Investigation, Writing--original draft, Writing--review and editing.
Viet Bui: Investigation, Writing--original draft, Writing--review and editing.
Tien Mai: Conceptualization, Investigation, Supervision, Writing--original draft,
Writing--review and editing.

\section*{Funding}
This research did not receive any specific grant from funding agencies in the
public, commercial, or not-for-profit sectors.

\section*{Data Availability}
No new data were created or analyzed in this study.

\section*{Declaration of Competing Interest}
The authors declare that they have no known competing financial interests or
personal relationships that could have appeared to influence the work reported
in this paper.

\clearpage
\phantomsection
\addcontentsline{toc}{section}{References}
\bibliographystyle{elsarticle-harv}
\bibliography{references}
\end{document}